\pdfoutput=1
\relax
\documentclass[letterpaper]{article} 
\usepackage{aaai21}  
\usepackage{times}  
\usepackage{helvet} 
\usepackage{courier}  
\usepackage[hyphens]{url}  
\usepackage{graphicx} 
\usepackage{natbib}  
\usepackage{caption} 
\usepackage[boxed,vlined]{algorithm2e}
\usepackage{algpseudocode}
\usepackage{paralist}
\usepackage{bbding}
\usepackage{diagbox}
\usepackage{url}
\usepackage{caption}
\usepackage{subcaption}
\usepackage[table]{xcolor}
\usepackage{amsmath}
\usepackage{multirow}
\usepackage{adjustbox}
\usepackage[switch]{lineno}
\usepackage{tikz}

\newtheorem{observation}{Observation}

\newcommand{\method}{\textsc{CGEM}\xspace} 

\title{\textit{Contact Graph Epidemic Modelling} of COVID-19 for Transmission and Intervention Strategies}

\author{Abby Leung, Xiaoye Ding, Shenyang Huang, Reihaneh Rabbany\\}
\affiliations{
    School of Computer Science, McGill University \\
     Mila, Quebec Artificial Intelligence Institute\\
    \{oi.k.leung, xiaoye.ding, shenyang.huang, reihaneh.rabbany\}@mail.mcgill.ca
}

\newcommand{\covid}{COVID-19 }

\begin{document}
\maketitle
\begin{abstract}
The coronavirus disease 2019 (COVID-19) pandemic has quickly become a \textit{global} public health crisis unseen in recent years. It is known that the structure of the human contact network plays an important role in the spread of transmissible diseases. In this work, we study a structure aware model of COVID-19 (\method). This model becomes similar to the classical compartment-based models in epidemiology if we assume the contact network is a Erd\H{o}s-Re\'nyi (ER) graph, i.e. everyone comes into contact with everyone else with the same probability. In contrast, \method is more expressive and allows for plugging in the actual contact networks, or more realistic proxies for it. Moreover, \method enables more precise modelling of \textit{enforcing} and \textit{releasing} different non-pharmaceutical intervention (NPI) strategies.
Through a set of extensive experiments, we demonstrate significant differences between the epidemic curves when assuming different underlying structures. More specifically we demonstrate that the compartment-based models are overestimating the spread of the infection by a factor of 3, and under some realistic assumptions on the compliance factor, underestimating the effectiveness of some of NPIs,  mischaracterizing others (e.g. predicting a later peak), and underestimating the scale of the second peak after reopening. 




\end{abstract}

\section{Introduction}
\label{sec:intro}







\begin{figure}[t]
    \begin{center}
        \centerline{\includegraphics[width=0.9\columnwidth]{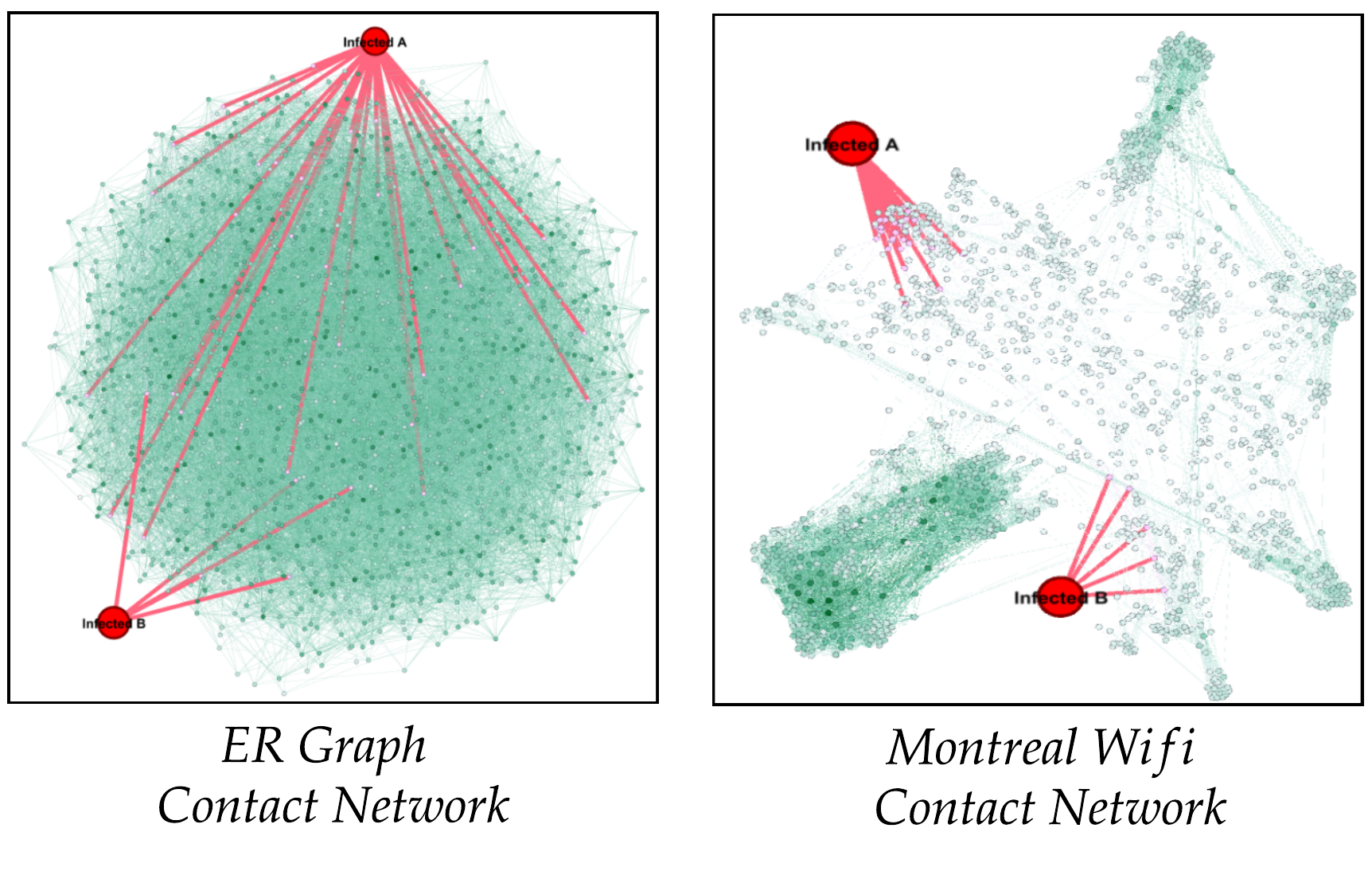}}
        \caption{\textbf{\method can utilize realistic contact networks}. Standard compartment models assume an underlying ER contact network, whereas real networks have a non-random structure as seen in Montreal Wifi example. In each network, two infected patients with 5 and 29 edges are selected randomly and the networks in comparison have the same number of nodes and edges. In Wifi network, infected patients are highly likely to spread their infection in their local communities while in ER graph they have a wide-spread reach.}
        \label{Fig:compare}
    \end{center}
    \vskip -0.3in
\end{figure} 

Epidemic modelling of \covid has been used to inform public health officials across the globe and the subsequent decisions have significantly affected every aspect of our lives, from financial burdens of closing down businesses and the overall economical crisis, to long term affect of delayed education, and adverse effects of confinement on mental health. Given the huge and long-term impact of these models on almost everyone in the world, it is crucial to design models that are as realistic as possible to correctly assess the cost benefits of different intervention strategies. Yet, current models used in practice have many known issues. In particular, the commonly-used compartment based models from classical epidemiology do not consider the structure of the real world contact networks. It has been shown previously that contact network structure changes the course of an infection spread significantly~\cite{keeling2005implications,bansal2007individual}. 
In this paper, we demonstrate the structural effect of different underlying contact networks in COVID-19 modelling. 

Non-pharmaceutical Interventions~(NPIs) played a significant role in limiting the spread of COVID-19. Understanding effectiveness of NPIs is crucial for more informed policy making at public agencies~(see the timeline of NPIs applied in Canada in Table~\ref{tab:NPIs}). However, the commonly used compartment based models are not expressive enough to directly study different NPIs. For example, \citet{ogden2020predictive} described the predictive modelling efforts for COVID-19 within the Public Health Agency of Canada. To study the impact of different NPIs, they used an agent-based model in addition to a separate deterministic compartment model. One significant disadvantage of the compartment model is its inability to realistically model the closure of public places such as schools and universities. This is due to the fact that compartment models assume that each individual has the same probability to be in contact with every other individual in the population which is rarely true in reality. Only by incorporating real world contact networks into compartment models, one can disconnect network hubs to realistically simulate the effect of closure. Therefore, \citet{ogden2020predictive} need to rely on a separate stochastic agent-based model to model the closure of public places. In contrast, our proposed \method is able to directly model all NPIs used in practice realistically.
 
In this work, we propose to incorporate structural information of contact network between individuals and show the effects of NPIs applied on different categories of contact networks. In this way, we can 1) more realistically model various NPIs, 2) avoid the imposed homogeneous mixing assumption from compartment models and utilize different networks for different population demographics. First, we perform simulations on various synthetic and real world networks to compare the impact of the contact network structure on the spread of disease. Second, we demonstrate that the degree of effectiveness of NPIs can vary drastically depending on the underlying structure of the contact network. We focus on the effects of 4 widely adopted NPIs: 1) quarantining infected and exposed individuals, 2) social distancing, 3) closing down of non-essential work places and schools, and 4) the use of face masks. Lastly, we simulate the effect of re-opening strategies and show that the outcome will depend again on the assumed underlying structure of the contact networks.

To design a realistic model of the spread of the pandemic, we also used a wifi hotspot network from Montreal to simulate real world contact networks. Given our data is from Montreal, we focus on studying Montreal timeline but the basic principles are valid generally and \method is designed to be used with any realistic contact network. We believe that \method can improve our understanding on the current COVID-19 pandemic and be informative for public agencies on future NPI decisions. 


\textbf{Summary of contributions}: 
\begin{itemize}
    \item We show that structure of the contact networks significantly changes the epidemic curves and the current compartment based models are subject to overestimating the scale of the spread 
    \item We demonstrate the degree of effectiveness of different NPIs depends on the assumed underlying structure of the contact networks  
    \item We simulate the effect of re-opening strategies and show that the outcome will depend again on the assumed underlying structure of the contact networks 
\end{itemize}
\textbf{Reproducibility}: Code for the model and synthetic network generation are in supplementary material. The real-world data can be accessed through the original source. 

\begin{table}[t]
    \centering
    \footnotesize
    \begin{tabular}{l|ll}
        Date & Location & Event(s)\\
        \hline
        Mar. 11 & [Worldwide] & WHO declares global pandemic \\
        Mar. 12 & [QC] & returning travellers to self-isolate \\
        & [ON] & close public schools \\
        Mar. 13 & [ON, AB] & cancel events $> 250$ \\
        & [BC, MB] & cancel events $> 250$ \\
        & [NS, NB] & discourage gatherings $> 150$ \\
        Mar. 14 & [QC, ON] & ban visits to long term care facilities \\
        Mar. 15 & [NS] & close schools, childcare, casinos \\
        & & ban visits to long term care facilities \\
        && ban gatherings over 150 \\
        Mar. 16 & [Canada] & close borders, excluding US. \\ 
        & [Canada] & mandatory 14 days quarantine \\
        & [QC] & close schools, universities, and daycares\\
        Mar. 17 & [ON, AB] & ban public events of over 50 \\
        & [BC] & close schools, restaurants, and bars \\
        Mar. 19 & [NB] & close most businesses, gatherings $\leq 10$ \\
        Mar. 20 & [Canada] & close boarder with US \\
        Mar. 23 & [NS] & quarantine for domestic travellers \\
        & [Canada] & social distancing enforced \\
        & [ON, QC] & close all non-essential workplace \\
        Apr. 6 & [Canada] & advise to wear masks\\
        May 22 & [MTL] & allow outdoor gatherings $\leq 50$ \\
        & & ease social distancing for some \\
        May 25 & [MTL] & reopen shops with exterior entrance \\
        & [QC] & reopen manufacturers without restrictions \\ 
        June 15 & [MTL] & reopen personal and aesthetic care \\
        June 22 & [MTL] & reopen restaurants \\
        June 28 & [MTL] & reopen educational childcare facilities \\
        July 18 & [QC] & reopen offices \\
        Aug. 1 & [QC] & allow indoor gathering $\leq 250$\\
        Aug. 5 & [QC] & allow outdoor gathering $\leq 250$\\
        \hline
    \end{tabular}
    \caption{Timeline of Canada \covid selected NPI events based on~\cite{npi_timeline,reopen_timeline}}
    \label{tab:timeline}
    \vspace{-10pt}
\end{table}

\section{Related Work}
\label{sec:background}


\subsection{Network Structures and Disease Modelling}

 Different approaches have accounted for network structures in epidemiological modelling. Degree block approximation~\cite{barabasi2016network} considers the degree distribution of the network by grouping nodes with the same degree into the same block and assuming that they have the same behavior. Percolation theory methods~\cite{newman2002spread} can approximate the final size of the epidemic for networks with specified degree distributions. Recently, \citet{sambaturu2020designing} studied the EPICONTROL problem which aims to design effective vaccination strategies based on real and diverse contact networks. Various modifications are made to the compartment differential equations to account for the network effect~\cite{aparicio2007building, keeling2005implications, bansal2007individual}. Simulation-based approaches are often used when the underlying networks are complex and mathematically intractable. \citet{grefenstette2013fred} employed an agent-based model to simulate the dynamics of the SEIR model with a census-based synthetic population. The contact networks are implied by the behavior patterns of the agents. \citet{chen2020time} adopted the Independent Cascade (IC) model \cite{saito2008prediction} to simulate the disease propagation and used Facebook network as a proxy for the contact network. Social networks, however, are not always a good approximation for the physical contact networks. In our study, we attempt to better ground the simulations by inferring the contact networks from wifi hub connection records.

\subsection{Modelling Non-pharmaceutical Interventions}

Table~\ref{tab:NPIs} shows the simulation of different NPIs in literature when compared to our model. 
\citet{tuite2020mathematical} used an age-structured SEIR model to estimate the spread of COVID-19 in Ontario, Canada. When studying the effect of NPIs, their key outputs included final epidemic attack rates~(\% of population infected at the end of the 2-year period), prevalence of hospital admissions and ICU use, and death. They assumed the effect of physical-distancing measures were to reduce the number of contacts per day across the entire population. In addition, enhanced testing and contact tracing were assumed to move individuals with nonsevere symptoms from the infectious to isolated compartments. In this work, we also examine the effect of closure of public places which is difficult to simulate in a realistic manner for standard compartment models.

\begin{table}[t]
\begin{center}
\footnotesize
\begin{tabular}{l| c c c c c c}
\diagbox{Disease Models}{NPIs} & \rotatebox{90}{Quarantine} & \rotatebox{90}{Social Distancing} & \rotatebox{90}{Wearing Masks} & \rotatebox{90}{Closure} \\
\hline
\hline
\rowcolor{gray!20}\method~(ours) & \CheckmarkBold & \CheckmarkBold  & \CheckmarkBold  & \CheckmarkBold \\
\citet{tuite2020mathematical} & \CheckmarkBold     & \CheckmarkBold  &                 & \\
SEIR~\cite{ogden2020predictive} & \CheckmarkBold  & \CheckmarkBold  &                 &  \\
Agent-based~\cite{ogden2020predictive} &                &                 &                 & \CheckmarkBold \\
\citet{chen2020time} &                             & \CheckmarkBold  &                 & \CheckmarkBold  \\
\citet{reich2020modeling} & \CheckmarkBold         & \CheckmarkBold  &                 &  \\
\citet{ferguson2020report} & \CheckmarkBold        & \CheckmarkBold  &                 & \CheckmarkBold \\
\hline
\hline
\end{tabular}
\caption{\textbf{\method can realistically model all NPIs used in practice} while existing models miss one or more NPIs}
\label{tab:NPIs}
\end{center}
\end{table}

\citet{ogden2020predictive} described the predictive modelling efforts for COVID-19 within the Public Health Agency of Canada. They estimated that more than 70\% of the Canadian Population may be infected by COVID-19 if no intervention is taken. They proposed an agent-based model and a deterministic compartment model. In the compartment model, similar to \citet{tuite2020mathematical}, effects of physical distancing are modelled by reducing daily per capita contact rates. The agent model is used to separately simulate the effects of closing schools, workplaces and other public places. In this work, we compare the effects all NPIs used in practice through a unified model and show how different contact networks change the outcome of NPIs. 
In addition, \citet{ferguson2020report} employed an individual-based simulation model to evaluate the impact of NPIs, such as quarantine, social distancing and school closure. The number of deaths and ICU bed demand are used as proxies to compare the effectiveness of NPIs. In comparison, our model can directly utilize contact networks and we also model the impact of wearing masks. 
\citet{block2020social} proposed three selective social distancing strategies based on the observations that epidemic dynamics depends on the network structure. The strategies aim to increase network clustering and eliminate shortcuts and are shown to be more effective than naive social distancing. 
\citet{reich2020modeling} proposed a selective social distancing strategy which lower the mean degree of the network by limiting super-spreaders. The authors also compared the impact of various NPIs, including testing, contact tracing, quarantine and social distancing. 
Neural network based approaches~\cite{soures2020sirnet, dandekar2020neural} are also proposed to estimate the effectiveness of quarantine and forecast the spread of the disease.

\section{Methodology}
\label{sec:meth}






\newcommand{\placefi}[1]{\resizebox{0.18\textwidth}{!}{\input{fig/#1}}}

\newcommand{\ccg}{Contact Graph SEIR }
\subsection{\ccg}
In a classic SEIR model, referred to as base SEIR, the dynamics of the system at each time step can be described by the following equations~\cite{ARON1984665}:

\begin{minipage}{0.45\columnwidth}
{\hspace{30pt}
\small \begin{align*}
    &\frac{dS}{dt} = - \frac{\beta SI}{N}\\
   &\frac{dE}{dt} = \frac{\beta SI}{N} - \sigma E\\
    &\frac{dI}{dt} = \sigma E - \gamma I\\
    &\frac{dR}{dt} = \gamma I
\end{align*}
}%
\end{minipage}
\begin{minipage}{0.45\columnwidth}
\hspace{0.2\columnwidth}
\resizebox{0.18\textwidth}{!}{



\tikzset{every picture/.style={line width=0.75pt}} 

\begin{tikzpicture}[x=0.75pt,y=0.75pt,yscale=-1,xscale=1,baseline=-10mm]

\draw  [fill={rgb, 255:red, 184; green, 233; blue, 134 }  ,fill opacity=0.61 ] (269.36,65.71) -- (297.3,65.71) -- (297.3,90.04) -- (269.36,90.04) -- cycle ;
\draw  [fill={rgb, 255:red, 251; green, 233; blue, 18 }  ,fill opacity=0.37 ] (269.45,107.52) -- (297.4,107.52) -- (297.4,131.84) -- (269.45,131.84) -- cycle ;

\draw  [fill={rgb, 255:red, 244; green, 139; blue, 139 }  ,fill opacity=0.67 ] (270.08,149.63) -- (298.02,149.63) -- (298.02,173.95) -- (270.08,173.95) -- cycle ;

\draw  [fill={rgb, 255:red, 156; green, 136; blue, 118 }  ,fill opacity=0.41 ] (270.18,191.44) -- (298.12,191.44) -- (298.12,215.76) -- (270.18,215.76) -- cycle ;

\draw    (282.48,91.85) -- (282.48,104.39) ;
\draw [shift={(282.48,106.39)}, rotate = 270] [color={rgb, 255:red, 0; green, 0; blue, 0 }  ][line width=0.75]    (10.93,-3.29) .. controls (6.95,-1.4) and (3.31,-0.3) .. (0,0) .. controls (3.31,0.3) and (6.95,1.4) .. (10.93,3.29)   ;
\draw    (282.48,132.75) -- (282.48,145.29) ;
\draw [shift={(282.48,147.29)}, rotate = 270] [color={rgb, 255:red, 0; green, 0; blue, 0 }  ][line width=0.75]    (10.93,-3.29) .. controls (6.95,-1.4) and (3.31,-0.3) .. (0,0) .. controls (3.31,0.3) and (6.95,1.4) .. (10.93,3.29)   ;
\draw    (283.41,174.56) -- (283.41,187.1) ;
\draw [shift={(283.41,189.1)}, rotate = 270] [color={rgb, 255:red, 0; green, 0; blue, 0 }  ][line width=0.75]    (10.93,-3.29) .. controls (6.95,-1.4) and (3.31,-0.3) .. (0,0) .. controls (3.31,0.3) and (6.95,1.4) .. (10.93,3.29)   ;

\draw (277.69,71) node [anchor=north west][inner sep=0.75pt]    {$S$};
\draw (286.04,90) node [anchor=north west][inner sep=0.75pt]    {$\beta $};
\draw (286.04,138) node [anchor=north west][inner sep=0.75pt]    {$\sigma $};
\draw (286.04,177) node [anchor=north west][inner sep=0.75pt]    {$\gamma $};
\draw (275.84,198) node [anchor=north west][inner sep=0.75pt]    {$R$};
\draw (275.84,156) node [anchor=north west][inner sep=0.75pt]    {$I$};
\draw (275.84,112) node [anchor=north west][inner sep=0.75pt]    {$E$};

\end{tikzpicture}
}\vspace{10pt}\\
\end{minipage}

 where an individual can be in one of the 4 states: ($S$) susceptible, ($E$) exposed, ($I$) infected and can infect nodes that are susceptible, and ($R$) recovered at any given time step $t$. $\beta, \sigma, \gamma$ are the transition rates from $S$ to $E$, $E$ to $I$, and $I$ to $R$ respectively.

Similarly, in \method, an individual can be either $S$ susceptible, $E$ exposed, $I$ infected or $R$ recovered. We do not consider reinfection, but extensions are straightforward. Unlike the equation-based SEIR model which assumes homogeneous mixing, \method takes into account the contact patterns between the individuals by simulating the spread of a disease over a contact network. Each individual becomes a node in the network and the edges represent the connections between people.

Algorithm~\ref{alg:simulation} shows the pseudo code for \method \footnote{For brevity, we use the same notation for both set and size of set, i.e. here $S$ represents the number of susceptible in SEIR model, whereas in \method $S$ represents the set of susceptible individuals.}.
Given a contact network, we assume that a node comes into contact with all its neighbours at each time step. More specifically, at each time step, the susceptible neighbours of infected individuals will become infected with a transmission probability $\phi$, and enter the exposed state (illustrated below). We randomly select exposed nodes to become infected with probability $\sigma$ and let them recover with a probability $\gamma$. 
\begin{minipage}{3.9\columnwidth}
\hspace{20pt}    
\resizebox{0.18\textwidth}{!}{

\tikzset{every picture/.style={line width=0.75pt}} 

\begin{tikzpicture}[x=0.75pt,y=0.75pt,yscale=-1,xscale=1]

\draw  [fill={rgb, 255:red, 251; green, 233; blue, 18 }  ,fill opacity=0.37 ] (354,96.21) .. controls (354,89.71) and (359.28,84.43) .. (365.79,84.43) .. controls (372.29,84.43) and (377.57,89.71) .. (377.57,96.21) .. controls (377.57,102.72) and (372.29,108) .. (365.79,108) .. controls (359.28,108) and (354,102.72) .. (354,96.21) -- cycle ;

\draw  [fill={rgb, 255:red, 184; green, 233; blue, 134 }  ,fill opacity=0.61 ] (520,100.21) .. controls (520,93.71) and (525.28,88.43) .. (531.79,88.43) .. controls (538.29,88.43) and (543.57,93.71) .. (543.57,100.21) .. controls (543.57,106.72) and (538.29,112) .. (531.79,112) .. controls (525.28,112) and (520,106.72) .. (520,100.21) -- cycle ;

\draw  [fill={rgb, 255:red, 244; green, 139; blue, 139 }  ,fill opacity=0.67 ] (427,149.21) .. controls (427,142.71) and (432.28,137.43) .. (438.79,137.43) .. controls (445.29,137.43) and (450.57,142.71) .. (450.57,149.21) .. controls (450.57,155.72) and (445.29,161) .. (438.79,161) .. controls (432.28,161) and (427,155.72) .. (427,149.21) -- cycle ;

\draw  [fill={rgb, 255:red, 156; green, 136; blue, 118 }  ,fill opacity=0.41 ] (481,183.21) .. controls (481,176.71) and (486.28,171.43) .. (492.79,171.43) .. controls (499.29,171.43) and (504.57,176.71) .. (504.57,183.21) .. controls (504.57,189.72) and (499.29,195) .. (492.79,195) .. controls (486.28,195) and (481,189.72) .. (481,183.21) -- cycle ;

\draw  [fill={rgb, 255:red, 251; green, 233; blue, 18 }  ,fill opacity=0.37 ] (523,155.21) .. controls (523,148.71) and (528.28,143.43) .. (534.79,143.43) .. controls (541.29,143.43) and (546.57,148.71) .. (546.57,155.21) .. controls (546.57,161.72) and (541.29,167) .. (534.79,167) .. controls (528.28,167) and (523,161.72) .. (523,155.21) -- cycle ;

\draw  [fill={rgb, 255:red, 251; green, 233; blue, 18 }  ,fill opacity=0.37 ] (370,153.21) .. controls (370,146.71) and (375.28,141.43) .. (381.79,141.43) .. controls (388.29,141.43) and (393.57,146.71) .. (393.57,153.21) .. controls (393.57,159.72) and (388.29,165) .. (381.79,165) .. controls (375.28,165) and (370,159.72) .. (370,153.21) -- cycle ;

\draw  [fill={rgb, 255:red, 184; green, 233; blue, 134 }  ,fill opacity=0.61 ] (425,197.21) .. controls (425,190.71) and (430.28,185.43) .. (436.79,185.43) .. controls (443.29,185.43) and (448.57,190.71) .. (448.57,197.21) .. controls (448.57,203.72) and (443.29,209) .. (436.79,209) .. controls (430.28,209) and (425,203.72) .. (425,197.21) -- cycle ;

\draw  [fill={rgb, 255:red, 184; green, 233; blue, 134 }  ,fill opacity=0.61 ] (572,83.21) .. controls (572,76.71) and (577.28,71.43) .. (583.79,71.43) .. controls (590.29,71.43) and (595.57,76.71) .. (595.57,83.21) .. controls (595.57,89.72) and (590.29,95) .. (583.79,95) .. controls (577.28,95) and (572,89.72) .. (572,83.21) -- cycle ;

\draw  [fill={rgb, 255:red, 184; green, 233; blue, 134 }  ,fill opacity=0.61 ] (464,113.21) .. controls (464,106.71) and (469.28,101.43) .. (475.79,101.43) .. controls (482.29,101.43) and (487.57,106.71) .. (487.57,113.21) .. controls (487.57,119.72) and (482.29,125) .. (475.79,125) .. controls (469.28,125) and (464,119.72) .. (464,113.21) -- cycle ;

\draw    (420.5,114.43) -- (433.5,139.43) ;
\draw    (467.5,121.43) -- (446.5,139.43) ;
\draw [shift={(457,130.43)}, rotate = 139.4] [fill={rgb, 255:red, 0; green, 0; blue, 0 }  ][line width=0.08]  [draw opacity=0] (8.93,-4.29) -- (0,0) -- (8.93,4.29) -- cycle    ;
\draw    (393.57,153.21) -- (427,149.21) ;
\draw    (520,100.21) -- (485.5,108.43) ;
\draw    (572,83.21) -- (541.5,94.43) ;
\draw    (583.79,95) -- (578.79,117.43) ;
\draw    (568.5,122.43) -- (540.5,107.43) ;
\draw    (464,113.21) -- (426.57,105.21) ;
\draw [shift={(445.29,109.21)}, rotate = 192.06] [fill={rgb, 255:red, 0; green, 0; blue, 0 }  ][line width=0.08]  [draw opacity=0] (8.93,-4.29) -- (0,0) -- (8.93,4.29) -- cycle    ;
\draw    (482,176.21) -- (447.5,156.43) ;
\draw    (448.57,197.21) -- (481,183.21) ;
\draw    (403,104.21) -- (377.57,96.21) ;
\draw  [fill={rgb, 255:red, 244; green, 139; blue, 139 }  ,fill opacity=0.67 ] (403,104.21) .. controls (403,97.71) and (408.28,92.43) .. (414.79,92.43) .. controls (421.29,92.43) and (426.57,97.71) .. (426.57,104.21) .. controls (426.57,110.72) and (421.29,116) .. (414.79,116) .. controls (408.28,116) and (403,110.72) .. (403,104.21) -- cycle ;

\draw  [fill={rgb, 255:red, 184; green, 233; blue, 134 }  ,fill opacity=0.61 ] (568,128.21) .. controls (568,121.71) and (573.28,116.43) .. (579.79,116.43) .. controls (586.29,116.43) and (591.57,121.71) .. (591.57,128.21) .. controls (591.57,134.72) and (586.29,140) .. (579.79,140) .. controls (573.28,140) and (568,134.72) .. (568,128.21) -- cycle ;

\draw    (436.79,185.43) -- (438.79,161) ;
\draw [shift={(437.79,173.21)}, rotate = 274.68] [fill={rgb, 255:red, 0; green, 0; blue, 0 }  ][line width=0.08]  [draw opacity=0] (8.93,-4.29) -- (0,0) -- (8.93,4.29) -- cycle    ;
\draw    (534.79,143.43) -- (531.79,112) ;

\draw (454+3,123+8) node [anchor=north west][inner sep=0.75pt]    {$\phi $};
\draw (438+3,85+8) node [anchor=north west][inner sep=0.75pt]    {$\phi $};
\draw (420+3,160+8) node [anchor=north west][inner sep=0.75pt]    {$\phi $};
\draw (571+3,116+8) node [anchor=north west][inner sep=0.75pt]    {$S$};
\draw (408+3,92+8) node [anchor=north west][inner sep=0.75pt]    {$I$};
\draw (467+3,101+8) node [anchor=north west][inner sep=0.75pt]    {$S$};
\draw (575+3,71+8) node [anchor=north west][inner sep=0.75pt]    {$S$};
\draw (428+3,185+8) node [anchor=north west][inner sep=0.75pt]    {$S$};
\draw (373+3,141+8) node [anchor=north west][inner sep=0.75pt]    {$E$};
\draw (526+3,143+8) node [anchor=north west][inner sep=0.75pt]    {$E$};
\draw (484+3,171+8) node [anchor=north west][inner sep=0.75pt]    {$R$};
\draw (432+3,137+8) node [anchor=north west][inner sep=0.75pt]    {$I$};
\draw (523+3,88+8) node [anchor=north west][inner sep=0.75pt]    {$S$};
\draw (357+3,84+8) node [anchor=north west][inner sep=0.75pt]    {$E$};

\end{tikzpicture}
}
\end{minipage}

\method offers the flexibility of accommodating different network structures and aligns better with the observation that people tend to interact with each other selectively and locally in the real world.


%


\begin{algorithm}[t]
\SetAlgoLined
\KwIn{Susceptible $S$, Exposed $E$, Infected $I$, Recovered $R$, transmission probability $\phi$, incubation rate $\sigma$, recovery rate $\gamma$}
\Repeat{end of simulation}{
    Contacts $\leftarrow$ []\;
    \For{i in I}{
        Contacts += neighbors(i)\;
    }
    $\Delta E$ $\leftarrow$ []\;
    \For{c in Contacts}{
        \If{ c in S and random(0, 1) $>$  $\phi$} {
            $\Delta E$ += c\;
        }
    }
    $\Delta I \leftarrow$ randomly choose $\sigma \times |E|$ nodes from $E$\; 
    $\Delta R$ $\leftarrow$ randomly choose $\gamma \times |I| $ nodes from $I$\;
    $I \leftarrow I + \Delta I - \Delta R$\;
    $R \leftarrow R + \Delta R$\;
}
\KwOut{$S, E, I, R$}
\caption{\method }
\label{alg:simulation}
\end{algorithm}

\subsection{Generalizing SEIR Model}
\method provides a generalized version of SEIR model which allows plugging in different structures. How can we draw connections between the parameters between the original model and our graph-based model? On the infection propagation process, the only parameter that differs from the original model is the transmission rate, $\phi$. Other parameters are directly matched with the original model. On the structure side, if substituting the ER graph with other synthetic graphs, e.g. BA model \cite{barabasi2016network}, the parameters of the synthetic graph generation could be adjusted to produce graphs with same sizes thus facilitating a fair comparison between different structures. We discuss details in the following sections.

\subsubsection{Inferring Transmission Rate}
By definition, $\beta$ represents the likelihood that a disease is transmitted from an infected to a susceptible in a unit time. \citet{barabasi2016network} assumes that on average each node comes into contact with $k$ neighbors, then the relationship between $\beta$ and the transmission rate $\phi$ can be expressed as: 
\begin{equation}
    \beta = \langle k \rangle \cdot \phi
    \label{eq:transmission}
\end{equation}
where $\langle k \rangle$ is the average degree of the nodes. 

In the case of a regular random network, all nodes have the same degree, i.e. $\langle k \rangle=k$ and equation~\ref{eq:transmission} can be reduced into: 
\begin{equation}
    \beta = k \cdot \phi
    \label{eq:transmission1}
\end{equation}

The homogeneous mixing assumption made by the standard SEIR model can be well simulated by running \method over a regular random network, we propose to bridge the two models with the following procedure: 
\begin{enumerate}
    \item Fit the classic SEIR model to real data to estimate $\beta$. 
    \item Run \method over regular random networks with different values of $k$ and with $\phi$ derived from equation~\ref{eq:transmission1}.
    \item Choose $k=k^{*}$ which produce the best fit to the predictions of the classic SEIR model.
\end{enumerate}

The regular random network with average degree $k^{*}$ would be the contact network the classic SEIR model is approximating and $\phi^{*} = \beta / k^{*}$ would be the implied transmission rate. We will use this transmission rate for other contact networks studied, so that the dynamics of the disease (transmissibility) is fixed and only the structure of contact graph changes.

\subsubsection{Tuning Synthetic Network generators}
As a proxy for actual contact networks which are often not available, we can pair \method with synthetic networks with more realistic properties,  comparable to real world networks e.g. heavy-tail degree distribution, small average shortest path, etc. 
To adjust the parameters of these generators, we can reframe the problem as: given transmission rate $\phi^{*}$ and population size $n$, are there other networks which can produce the same infection curve? 
For this, we can carry out similar procedures as above. For example, we can run \method with transmission rate $\phi^{*}$ over scale-free networks generated from different values of $m_{BA}$, where $m_{BA}$ is the number of edges a new node can form in the Barabasi Albert algorithm~\cite{barabasi2016network}. $m_{BA}$ which produces the best fit to the infection curve gives us a synthetic contact network that is realistic in terms of number of edges compared to the real contact network.

\subsection{Modelling NPIs with \ccg }
Here we explain how different NPIs can be modelled directly in \method as changes in the underlying structure. 
\subsubsection{Quarantine}
How can we model the quarantining and self-isolation of exposed and infected individuals? 
Exposed individuals have come into close contact with an infected person and are considered to have high risk of contracting. 
In an ideal world, most, if not all, infected individuals would be easily identifiable and quarantined. However, in reality, over 40\%~\cite{he2020temporal} of infected cases are asymptomatic and not all are identified immediately or at all and therefore can go on to infect others unintentionally. To account for this in our model, we apply quarantining by removing all edges from a subset of exposed and infected nodes.

\subsubsection{Social Distancing}
Social distancing reduces opportunities of close contacts between individuals by limiting contacts to those from the same household and staying at least 6 feet apart from others when out in public. 
In \method, a percentage of edges from each node are removed to simulate the effects of social distancing to different extent. 

\subsubsection{Wearing Masks}
Masks are shown to be effective in reducing the transmission rate of \covid with a relative risk (RR) of 0.608~\cite{Ollila2020.07.31.20166116}. We simulate this by assigning a mask wearing state to each node and varying the transmissibility, $\phi$, based on whether 2 nodes in contact are wearing masks or not. We define the new transmission rate with this NPI, $\phi_{mask}$ as follows:
\[
    \phi_{mask} = 
\begin{cases}
    m_2 \cdot \phi ,& \text{if both nodes wearing masks} \\
    m_1 \cdot \phi ,& \text{if 1 node wearing masks} \\
    m_0 \cdot \phi ,& \text{otherwise} \\
\end{cases}
\]

\subsubsection{Closure: Removing Hubs}
Places of mass gathering (e.g. schools and workplaces) put large number of people in close proximity. If infected individuals are present in these locations, they can have a large number of contacts and very quickly infect many others. 
In a network, these nodes with a high number of connections, or degree, are known as hubs. By removing the top degree hubs, we simulate the effects of cancelling mass gathering, and closing down schools and non-essential workplaces. In \method, we remove all edges from $r$\% of top degree nodes to simulate the closure of schools and non-essential workplaces. However, some hubs, such as (workers in) grocery stores and some government agencies,  must remain open, so we assign each hub a successful removal rate of $p_{success}$ to control this effect.

\subsubsection{Compliance}
Given the NPIs are complied by majority but not all the individuals, we randomly assign a fixed percentage of the nodes as non-compilers. We set this to 26\% in all the simulations based on a recent survey~\cite{ipsos}.

\subsection{Reopening Strategies}
Due to the economical and psychological impacts of a complete lockdown on the society, it is critical to know how safe it is to resume commercial and social activities once the pandemic has stabilized. Therefore, we also investigate the impact of relaxing each NPIs and the risk of a second wave infection. 
More specifically, we simulate a complete reversing of the NPIs, by adding back the edges that were removed when the NPI was applied at first, to return the underlying structure to its original form.  




\section{Experiments}
\label{sec:exp}
\subsection{Dataset description}
We compare the spread of \covid with synthetic and real world networks. These networks include 3 synthetic networks, (1) the Regular random network, where all nodes have the same degree,  (2) the Erd\H{o}s-Re\'nyi random network, where the degree distribution is Poisson distributed, (3) the Barabasi Albert network, where the degree distributions follows a power law. Additionally, we analyzed 4 real world network, the USC35 network from the Facebook100 dataset~\cite{TRAUD20124165}, consisting of Facebook friendship relationship links between students and staffs at the University of Southern California in September 2005, and 3 snapshots of a real world wifi hotspot network from Montreal~\cite{ilesansfil-wifidog-20151106}, a network often used as a proxy for human contact network while studying disease transmission~\cite{info:doi/10.2196/jmir.3720, yang2020targeted}. In the Montreal wifi network, edges are formed between nodes (mobile phones) that are connected to the same public wifi hub at the same time. 
As shown in Table~\ref{tab:graph_prop}, each of the 7 networks consist of ~17,800 nodes, consistent with 1/100th of the population of the city of Montreal, and have between ~110,000 to ~220,000 edges, with the exception of the USC network. Due to the aggregated nature of the USC dataset, edge sampling is enforced during the contact phase in order to obtain reasonable disease spread. The synthetic networks are in general more closely connected than the Montreal wifi networks, despite having similar number of nodes and edges. Only the largest connected component is considered in all networks. 

\begin{table}[h]
\footnotesize
    \begin{tabular}{l|c|c|c}
        \hline
         Network & \# nodes & \# edges & Avg. shortest path \\
        \hline
        Regular & 17,800 & 186,900 & ~3.6 \\
        Erdo Renyi & 17,800 & ~220,000 & ~3.4 \\
        Barabasi Albert & 17,800 & ~160,000 & ~3.2 \\
        USC & 17,444 & 801,853 & ~2.8 \\
        Wifi 1 & 17,844 & 115,064 & ~4.1 \\
        Wifi 2 & 17,841 & 111,760 & ~4.2 \\
        Wifi 3 & 17,889 & 176,893 & ~4.9 \\
    \end{tabular}
    \caption{Properties of the contact graphs studied }
    \label{tab:graph_prop}
\end{table}

All Montreal and Quebec infection numbers are obtained from Santé Montréal~\cite{mtl_infection} and the Institut national de santé publique du Québec~\cite{qc_infection} respectivly.

\begin{table}[htb]
    \small
    \begin{tabular}{l|l|l|l|l}
        \hline
        Model & Param & Type & Value & Description\\
        \hline
        \multirow{9}{*}{All} & S & $i$ & 17,796 & \# initial susceptible nodes \\
        & E & $i$ & 3 &  \# initial exposed nodes \\
        & I & $i$ & 1 &  \# initial infected nodes \\
        & R & $i$ & 0 &  \# initial recovered nodes \\
        & $\sigma$ & $c$ & 1/5 & transition rate $E \rightarrow I$ \\
        & $\gamma$ & $c$ & 1/14 & recovery rate \\
        & $n$ & $c$ & 17800 & number of nodes \\
        & $\phi$ & $f$ & 0.0371 & transmission probability \\
        & $\beta$ & $f$ & 0.78 & transition rate $S \rightarrow E$  \\
        Regular & $k$ & $f$ & 21 & degree of each node \\
        ER & $P_{ER}$ & $f$ & 0.0014 & probability of an edge \\
        BA & $m_{BA}$ & $f$ & 10 & \# edges on new node \\
        \hline
    \end{tabular}
    \caption{Parameters for \method, Type $i$, $c$, $f$ stands for initial,  fitted, and constant parameters respectively. The values for constant parameters are set based on ~\cite{whocjm}.}
    \label{tab:algorithm1_param}
\end{table}


\subsection{Structures Changes the Epidemic Curves}
The structure of the contact network plays an important role in the spread of a disease~\cite{bansal2007individual}. It dictates how likely susceptible nodes will come into contact with infected ones and therefore it is crucial to evaluate how the disease will spread on each network with the same initial parameters. 
Here, the classic SEIR model is fitted against the infection rates from the first of the 100th case in Montreal to April 4 to obtain $\beta$, which is before any NPI is applied. With Eq.~\ref{eq:transmission1}, the transmission rate, $\phi$, is estimated to be $0.0371$ and is used across all networks. 
In all experiments, we also seed the population with the same initial number of 3 exposed nodes and 1 infected node. The parameters used to generate synthetic networks are obtained following the procedures described in the previous session. All results are averaged across 10 runs. The grey shaded region shows the $95\%$ confidence interval of each curve. %

\begin{figure}[t]
     \centering
     \begin{subfigure}[b]{0.25\textwidth}
         {\centering
         \includegraphics[width=\textwidth]{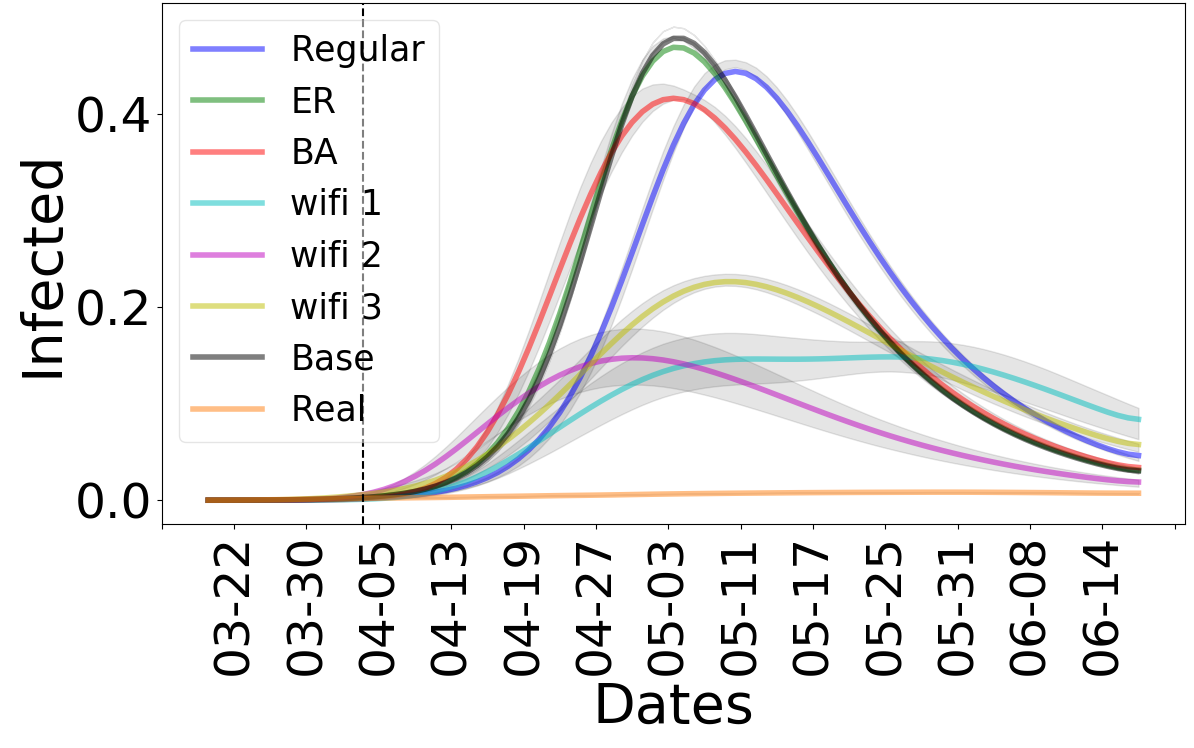}
         }
         \label{fig:nonpi}
     \end{subfigure}
     \begin{subfigure}[b]{0.25\textwidth}
         {\centering
         \includegraphics[width=\textwidth]{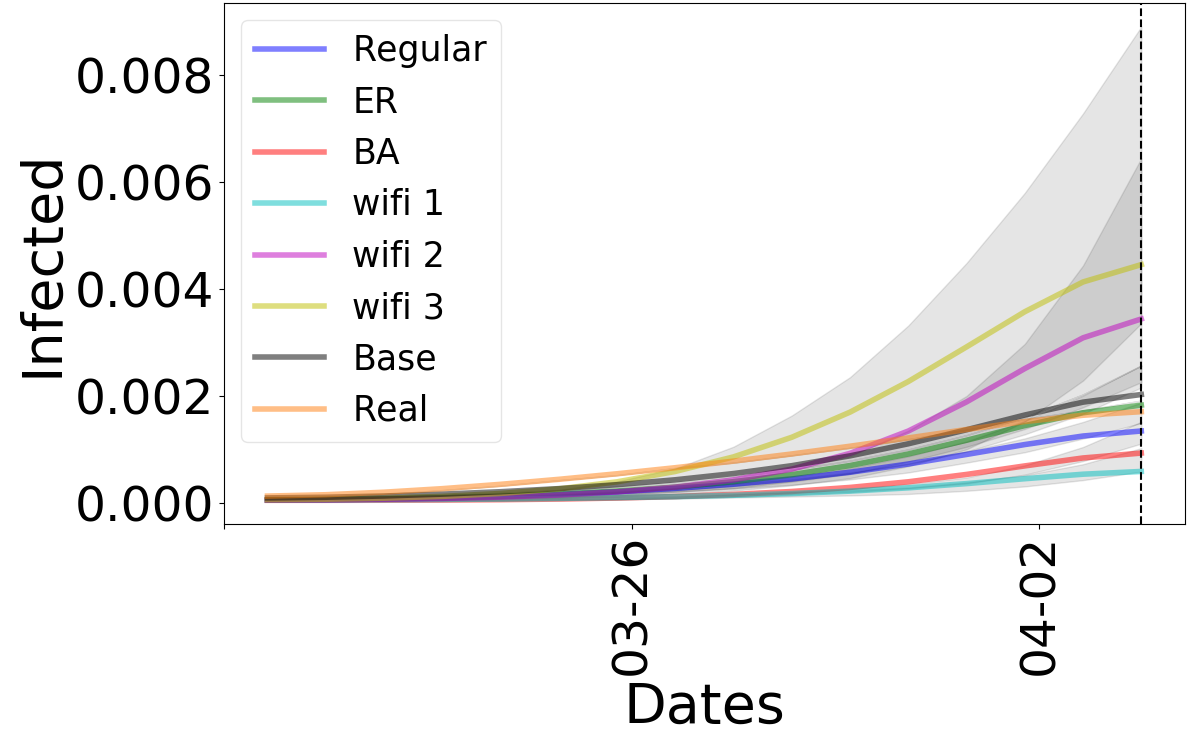}
         }
         \label{fig:nonpifirst16}
     \end{subfigure}
        \caption{(left) projects spread of the pandemic on all network structures without any NPI interventions. (right) spread of the pandemic during the time period fitted to the real data. The black dotted line indicates the date up to which the infection rate of the synthetic networks is fitted against the real world infection data, which coincides with the start of closed down, first NPI, in real-world. }
        \label{fig:nonpigraphs}
\end{figure}

As shown in Figure \ref{fig:nonpigraphs}, the ER network fits the base SEIR model almost perfectly--compare green 'ER' and black 'base' curves. 

\begin{observation}
    \method closely approximates  the base SEIR model when the contact network is assumed to be  Erd\H{o}s-Re\'nyi graph. 
\end{observation}

All networks drastically overestimates the spread of \covid when compared with real world data. This can be expected to some degree as in this experiment we are projecting the curves assuming no NPI is in effect which is not what happened in reality (see 'Real' orange curve). However, we observe that all 3 synthetic networks, including the ER model exceedingly overshoot, showing almost the entire population getting infected, whereas the real-world wifi networks predict a 3x lower peak. 

\begin{observation}
Assuming an Erd\H{o}s-Re\'nyi graph as the contact network overestimates the impact of \covid by more than a factor of 3 when compared with more realistic structures. 
\end{observation}


\subsection{Structure Changes the Effects of NPIs}

In order to limit the effects of the pandemic, the federal and provincial governments introduced a number of measures to reduce the spread of COVID-19. We simulate the effects of 4 different non-pharmaceutical interventions, or NPIs, at different strengths to determine their effectiveness. These include, (1) quarantining exposed and infected individuals, (2) social distancing between nodes, (3) removing hubs, and (4) the use of face masks.

\subsubsection{Quarantine}
We apply quarantining into our model on March 23. Where both Quebec and Canadian government have asked those who returned from foreign travels or experienced flu-like symptoms to self isolate. We remove all edges from 50, 75, and 95\% of exposed and infected nodes to simulate various strengths of quarantining. Figure~\ref{fig:quarantine} displays the effect of quarantining on different graph structures.
Quarantining infected and exposed nodes both reduces and delays the peak of all infection curve. 
However, the peak is not delayed as much in the wifi graphs as the ER graph predicts, which is important information in planning for the healthcare system. Out of all tested NPIs, applying quarantine has the most profound reduction on all infections curves.
\begin{observation}
Quarantining delays the peak of infection on the ER graph whereas the peak on the real world graphs are lowered but not delayed significantly. 
\label{obs:qdp}
\end{observation}
\begin{figure}[h]
     \centering
     \begin{subfigure}[b]{0.23\textwidth}
         \centering
         \includegraphics[width=\textwidth]{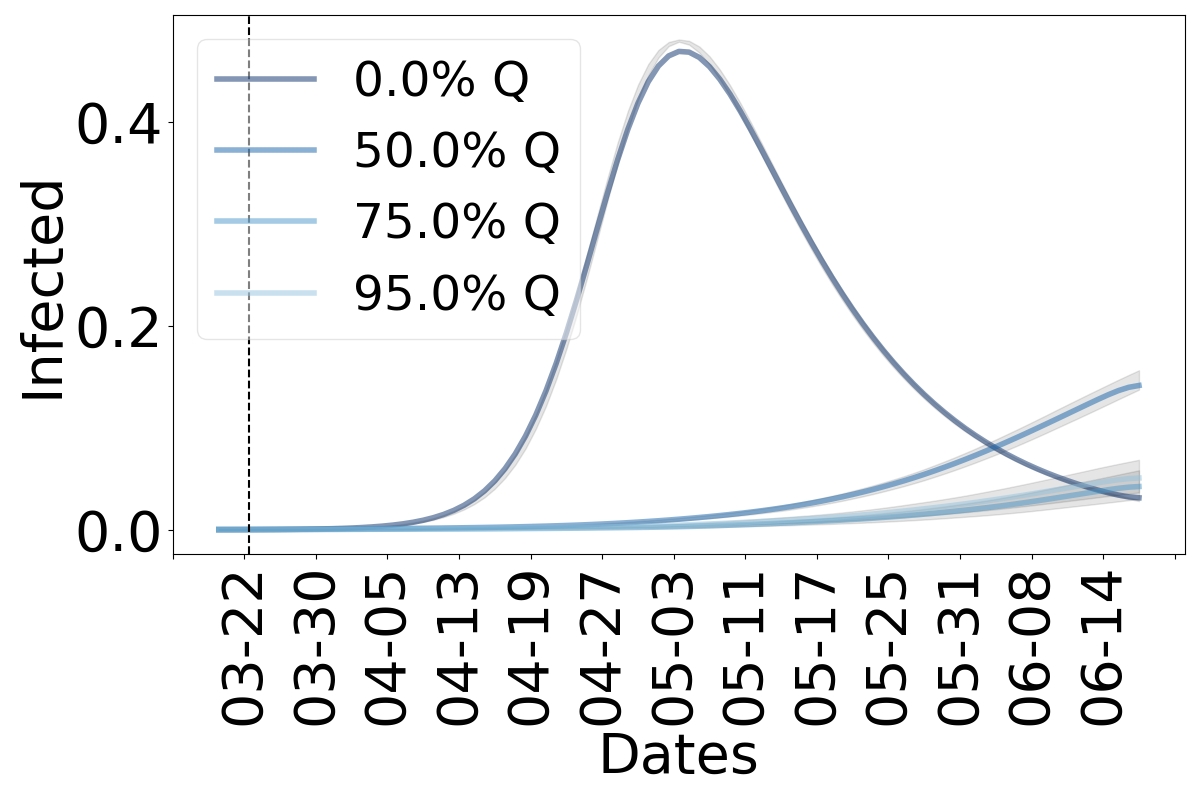}
         \caption{ER}
         \label{fig:quarantine er}
     \end{subfigure}
     \begin{subfigure}[b]{0.23\textwidth}
         \centering
         \includegraphics[width=\textwidth]{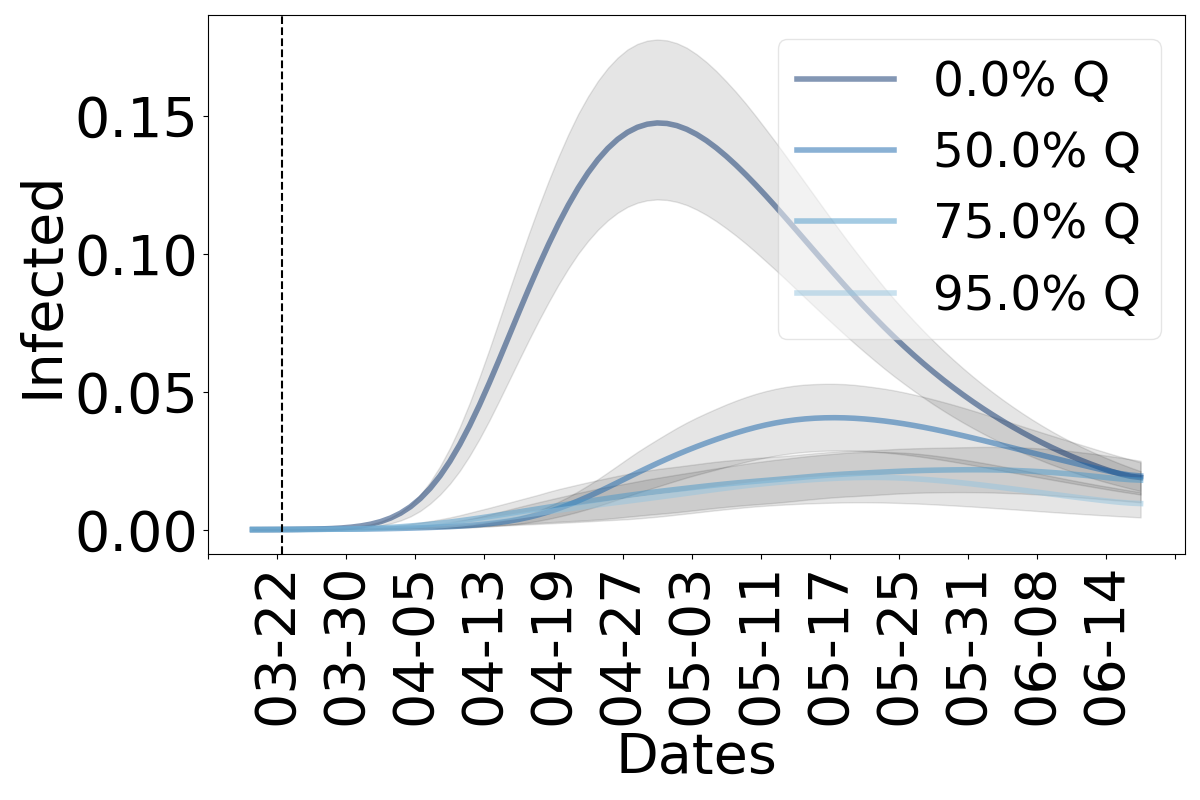}
         \caption{Wifi 2}
         \label{fig:quarantine wifi 2}
     \end{subfigure}
     \hfill
        \caption{Results of applying quarantine to 50, 75, and 95\% of nodes on ER and wifi2 graph structures. Other wifi networks show the same pattern and are omitted for brevity but are reported in the supplementary materials.} 
        \label{fig:quarantine}
\end{figure}

\subsubsection{Social Distancing}
reduces the number of close contacts. Different degrees of $10\%$, $30\%$, and $50\%$ of edges from each node is removed to simulate this. 
Figure~\ref{fig:Social distance} shows the effects of social distancing on the infection curves of each network structures. It is effective in reducing the peak of the pandemic on all networks but again delays the peaks only on synthetic networks. Similar to Observation~\ref{obs:qdp}, we have:
\begin{observation}
 Social distancing delays the peak of infection on the ER graph whereas the peak on the real world graphs are lowered but not delayed significantly. 
\end{observation}
\begin{figure}[h]
     \centering
     \begin{subfigure}[b]{0.23\textwidth}
         \centering
         \includegraphics[width=\textwidth]{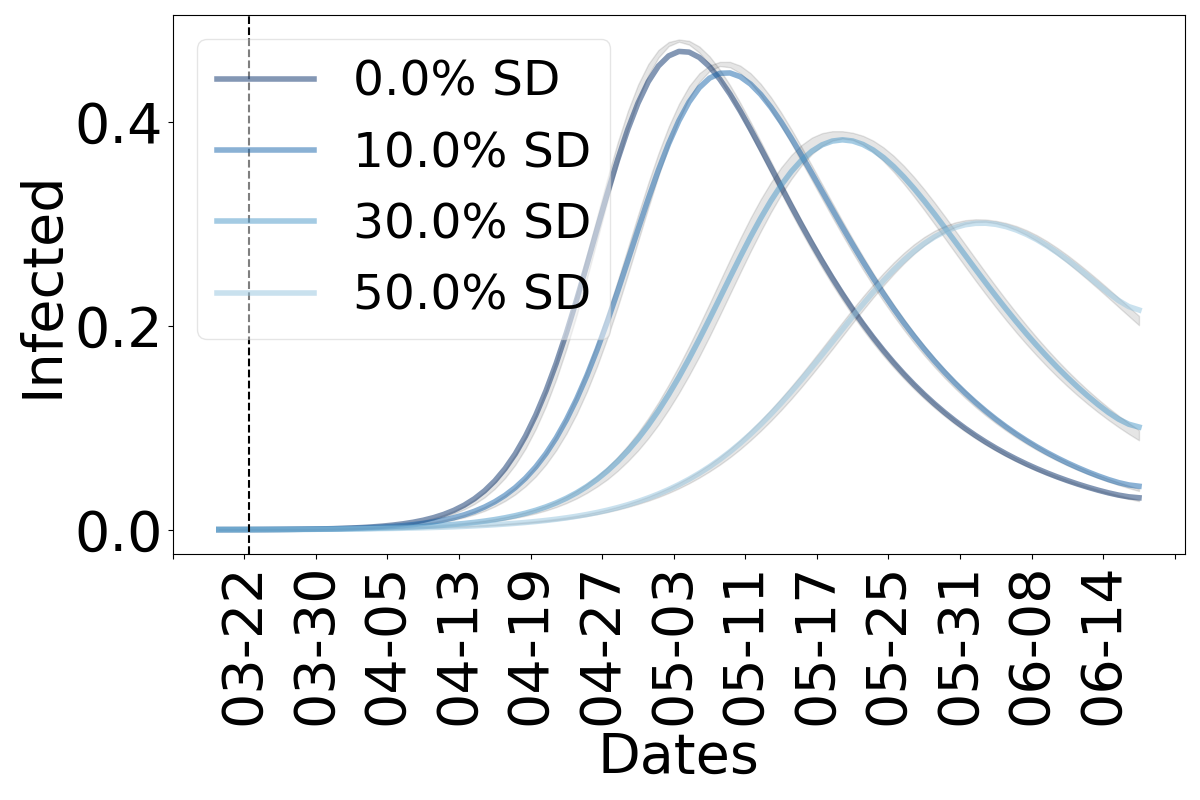}
         \caption{ER}
         \label{fig:Social distance er}
     \end{subfigure}
     \begin{subfigure}[b]{0.23\textwidth}
         \centering
         \includegraphics[width=\textwidth]{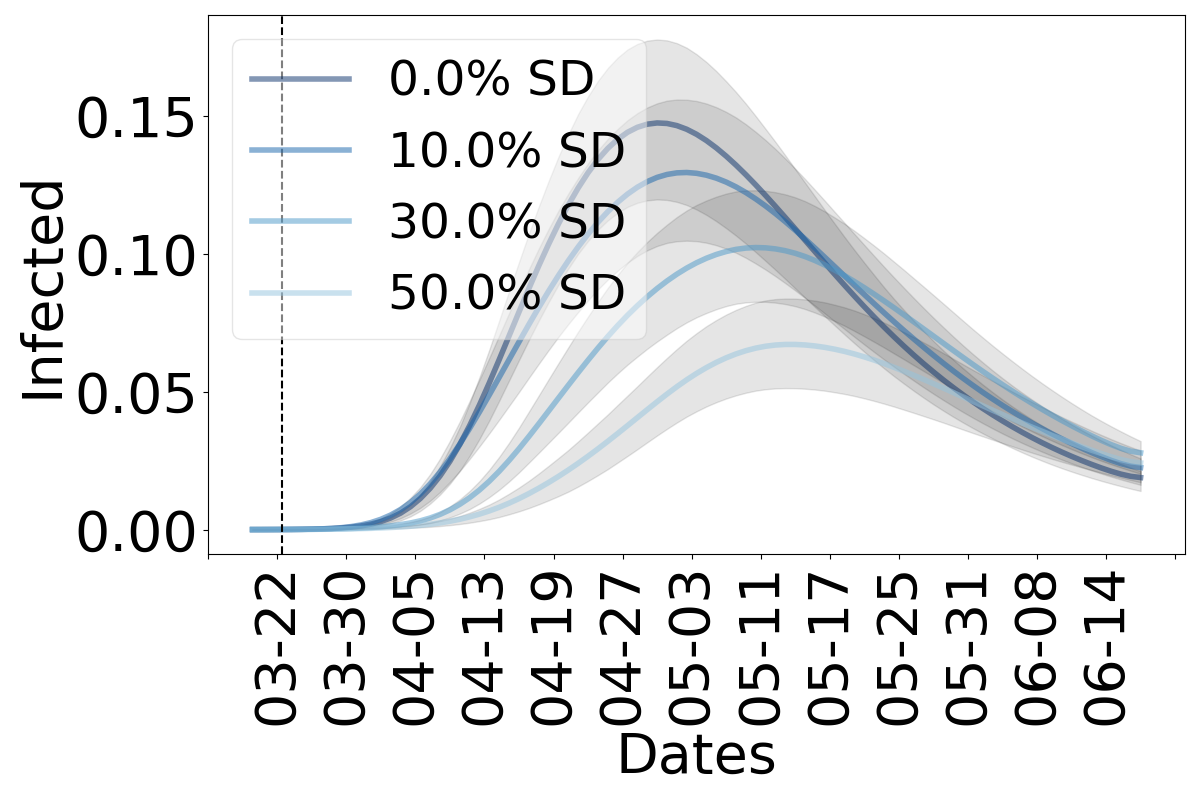}
         \caption{Wifi 2}
         \label{fig:Social distance wifi 2}
     \end{subfigure}
        \caption{Results of applying social distancing by removing 10, 30, and 50\% of edges to all nodes on ER and Wifi2 graph structures. Same pattern is observed for wifi1 and wifi3 (extended results are reported in the supplementary materials). 
        }
        \label{fig:Social distance}
\end{figure}

\subsubsection{Removing Hubs}
We remove all edges from 1\% of top degree nodes to simulate the closure of schools and 5 and 10\% of top degree nodes to simulate the closure of non-essential workplaces. These NPIs are applied on March 23 respectively, coinciding with the dates of school and non-essential business closure in Quebec. $p_{success}$ is set to 0.8 unless otherwise stated.
Figure~\ref{fig:hub} shows the effects of removing hubs. This NPI is very effective on the BA network and all 3 Montreal wifi networks since these networks have a power law degree distribution and hubs are present. However, it is not very effective on the regular and ER random networks. 

\begin{observation}
The ER graph significantly underestimates the effect of removing hubs. 
\end{observation}

\begin{figure}[h]
     \centering
     \begin{subfigure}[b]{0.23\textwidth}
         \centering
         \includegraphics[width=\textwidth]{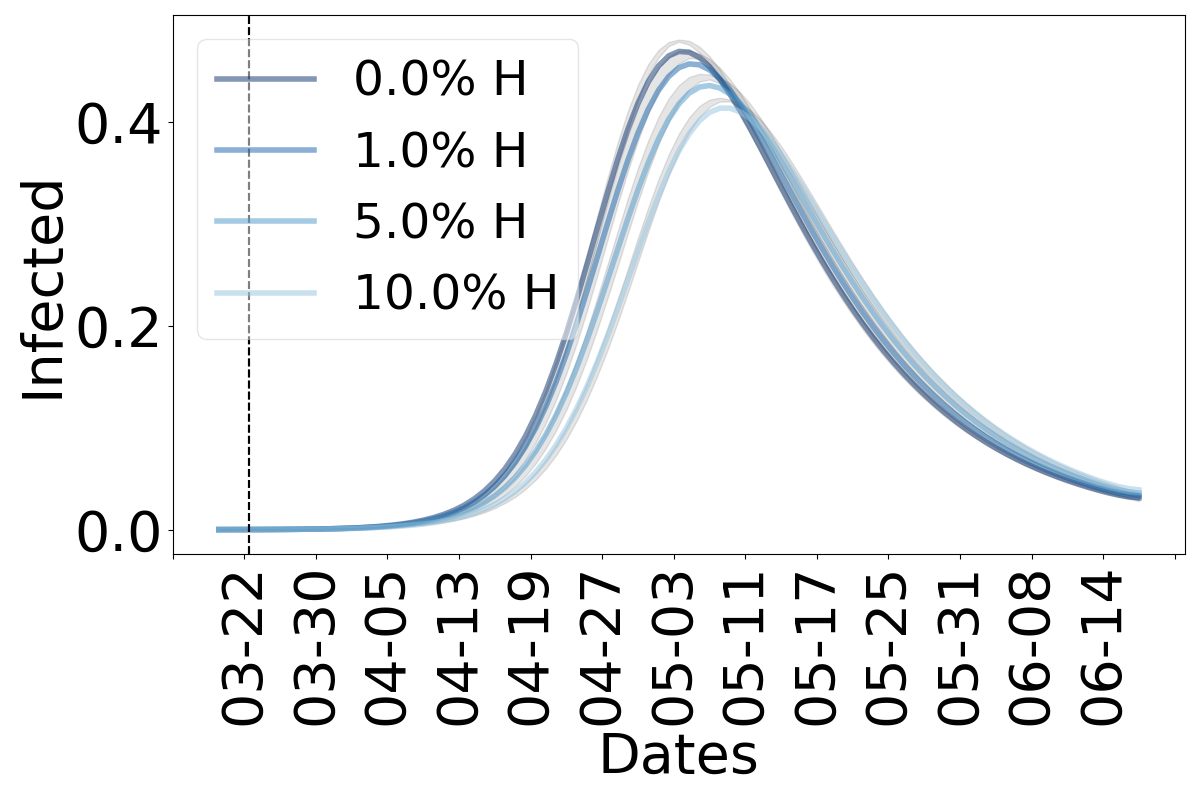}
         \caption{ER}
         \label{fig:hub er}
     \end{subfigure}
     \begin{subfigure}[b]{0.23\textwidth}
         \centering
         \includegraphics[width=\textwidth]{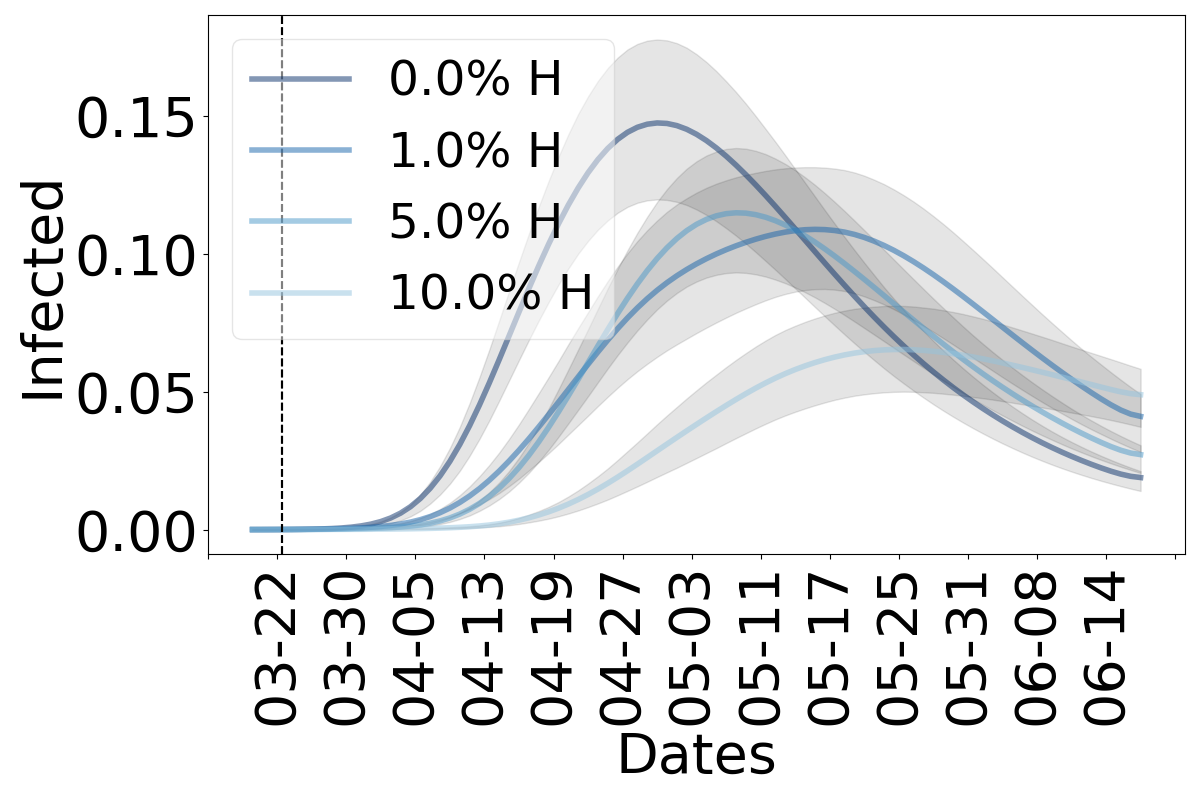}
         \caption{Wifi 2}
         \label{fig:hub wifi 2}
     \end{subfigure}
        \caption{Results of removing 1, 5, and 10\% of hubs from the ER and Wifi2 network. Again we see the same pattern on the other two Wifi graphs. }
        \label{fig:hub}
\end{figure}

Removing hubs is most effective on networks with a power law degree distribution since hubs act as super spreaders and removing them effectively contains the virus. However, no hubs are present in the ER and regular random network, and thus removing hubs reduces to removing random nodes. Luckily, real world contact networks have power law degree distributions, making a hubs removal an effective strategy in practice. 

\subsubsection{Wearing Masks}
we set $m_2=0.6$, $m_1=0.8$ and $m_0=1$, and use the following transmission rate, $\phi_{mask}$ in \method:
\[
    \phi_{mask} = 
\begin{cases}
    0.6 \cdot \phi ,& \text{if both nodes wearing masks} \\
    0.8 \cdot \phi ,& \text{if 1 node wearing masks} \\
    1 \cdot \phi ,& \text{otherwise} \\
\end{cases}
\]

Wearing masks is only able to flatten the infection curve on the synthetic networks but does not reduce the final epidemic attack rate, the total size of population infected, as shown in Figure~\ref{fig:mask}. However, in the real world wifi networks, wearing masks is able to both flatten the curve and also significantly reduce the final epidemic attack rate.

\begin{observation}
The ER graph significantly underestimates the effect of wearing masks in terms of the total decrease in the final attack rate. 
\end{observation}

\begin{figure}[h]
    \centering
    \includegraphics[width=0.42\textwidth]{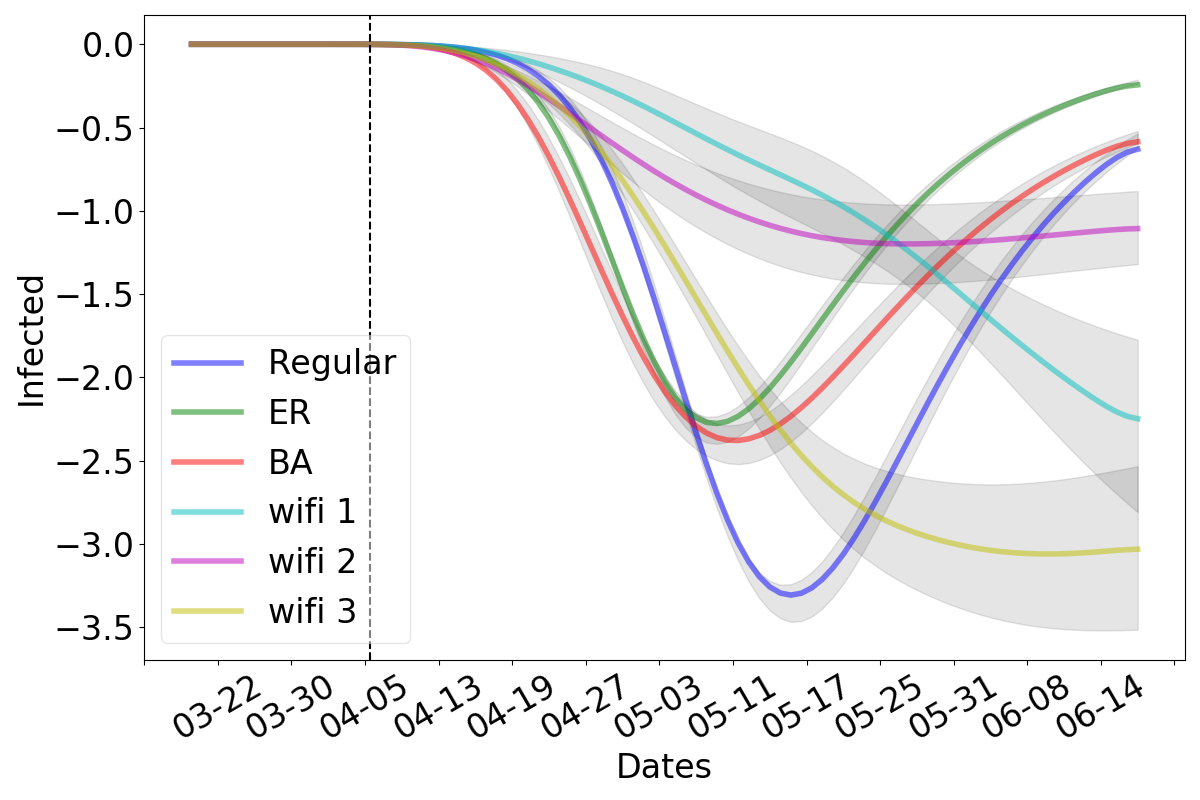}
    \caption{Difference between cumulative curves from wearing masks and not wearing masks. The cumulative curves represent the total impact, and the different shows how much drop  in final attack rate is estimated with the NPI enforced.}
    \label{fig:mask}
\end{figure}

\subsection{Structure Changes the Reopening Strategies}
We experiment with reopening of all the NPIs, but for brevity we only report the results for allowing hubs, which corresponds to the current reopening of schools and public places. The results form other NPIs are available in the extended results. 

\begin{figure}
    \centering
    \includegraphics[width=0.4\textwidth]{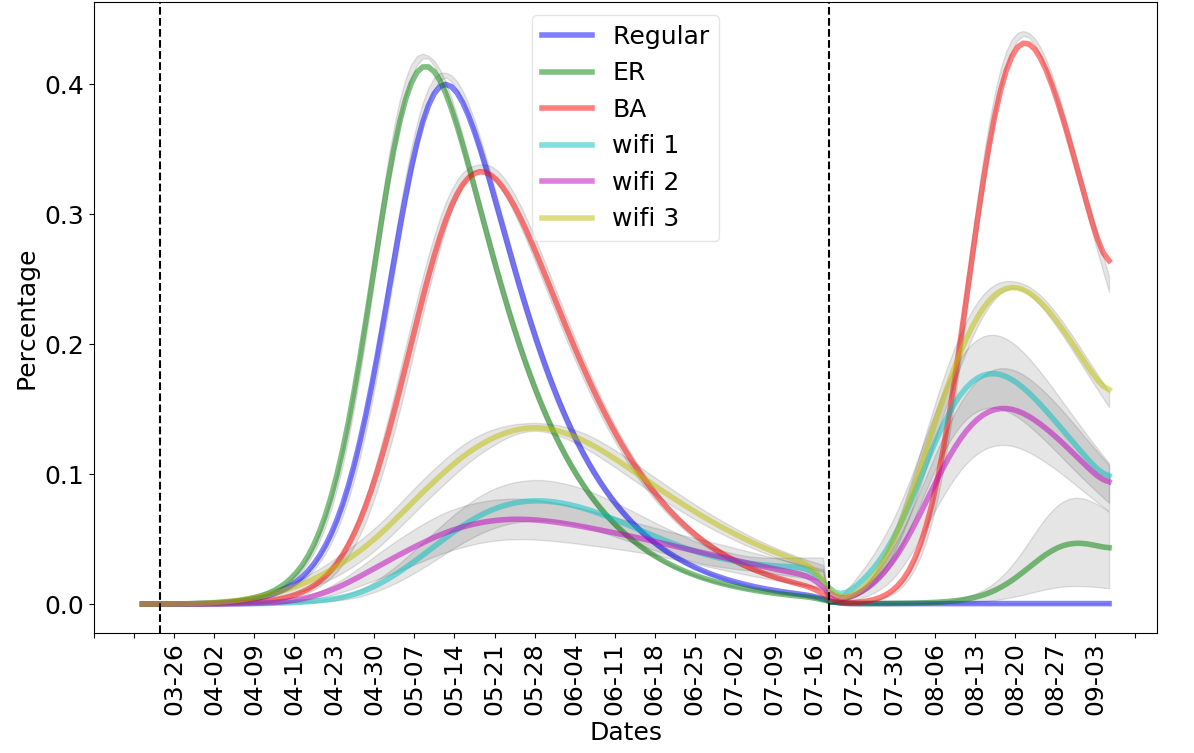}
    \caption{Reopening of 10\% of hubs removed earlier. The first vertical line represents when the NPI was enforced and the second vertical line shows when it is lifted. 
    }
    \label{fig:reopen}
\end{figure}

For removing hubs, we apply reopening on July 18 (denoted by the second vertical line in Figure~\ref{fig:reopen}), after many non-essential businesses and workplaces are allowed to open in Quebec. Because the synthetic networks estimates that most of the population would be infected before the hubs are reopened, we calibrate the number of infected and recovered individuals at the point of reopening to align with statistics available in the real world data. Therefore the simulation continues after reopening with all the models having the same number of susceptible individuals, otherwise int the ER graph, everyone is infected at that point. 
We can see in Figure~\ref{fig:reopen} that ER and regular random network significantly underestimates the extent of second wave infections. BA and the wifi networks all show second wave infections with a higher peak than the initial, prompting more caution when considering reopening businesses and schools.  


\begin{observation}
ER graph significantly underestimates the second peak after reopening public places, i.e. allowing back hubs. 
\end{observation}

\section{Conclusions}
\label{sec:concl}
In this paper, we propose to model \covid on contact networks (\method) and show that such modelling, when compared to traditional compartment based models, gives significantly different epidemic curves. Moreover, \method subsumes the traditional models while providing more expressive power to model the NPIs. We hope that \method could be used to achieve more informed policy making when studying reopening strategies for \covid.

\bibliography{ref}

\appendix

\section{Appendix}
\subsection{Network generation}
\subsubsection{Montreal wifi network}
3 snapshots of the Montreal wifi network are used in this paper with the following time periods: 2004-08-27 to 2006-11-30, 2007-07-01 to 2008-02-26, and 2009-12-02 to 2010-03-08. Each entry in the dataset consists of a unique connection id, a user id, node id (wifi hub), timestamp in, and timestamp out. Nodes in the network are the users in each connection. An edge forms between users who have connected to the same wifi hub at the same time. Connections are sampled with the aforementioned timestamp in dates to obtain $\sim$ 17800 nodes. Since there are many disconnected nodes in the wifi networks, only the giant connected component is used. 
\subsubsection{Synthetic networks}
We compared \method with the wifi networks with 3 synthetic network models, the regular, ER, and BA networks. In each of these models, we set the number of nodes to be 17,800 and fit respective parameters to best match the infection curve of the base model and the number of edges in the wifi networks. 
\begin{table}[h]
    \centering
    \begin{tabular}{c|c|c|c|c}
         \hline
         \multirow{ 2}{*}{Model} & \multirow{ 2}{*}{Param} & \multicolumn{2}{c|}{Range} & \multirow{ 2}{*}{best fit} \\
         & & begin & end & \\
         \hline
         Base SEIR & $\beta$ & 0.50 & 1.00 & 0.78 \\
         Regular & $k$ & 18 & 23 & 21 \\
         ER & $P_{ER}$ & 0.00130 & 0.00150 & 0.00140 \\
         BA & $m_{BA}$ & 5 & 15 & 10 \\
         \hline
    \end{tabular}
    \caption{Range of parameters fitted}
    \label{tab:param_range}
\end{table}
Table~\ref{tab:param_range} shows the range of numbers tested for each parameter and the fitted values. All parameters are fitted to the officially reported infected data and the mean squared error is minimized. 

\subsection{Computing Requirements}
All the experiments have been performed on a stock laptop.
\subsection{Assumptions}
The following assumptions are made in \method:
\begin{enumerate}
    \item Individuals who recover from \covid cannot be infected again 
    \item Symptomatic and asymptomatic individuals have the same transmission rate and they quarantine with the same probability
    \item A certain percentage of the population do not compile with NPIs regardless of their connection. 
\end{enumerate}


\subsection{Structure Changes the Effects of NPIs}
\subsubsection{Quarantine}
\begin{figure*}[h]
     \centering
     \begin{subfigure}[b]{0.33\textwidth}
         \centering
         \includegraphics[width=\textwidth]{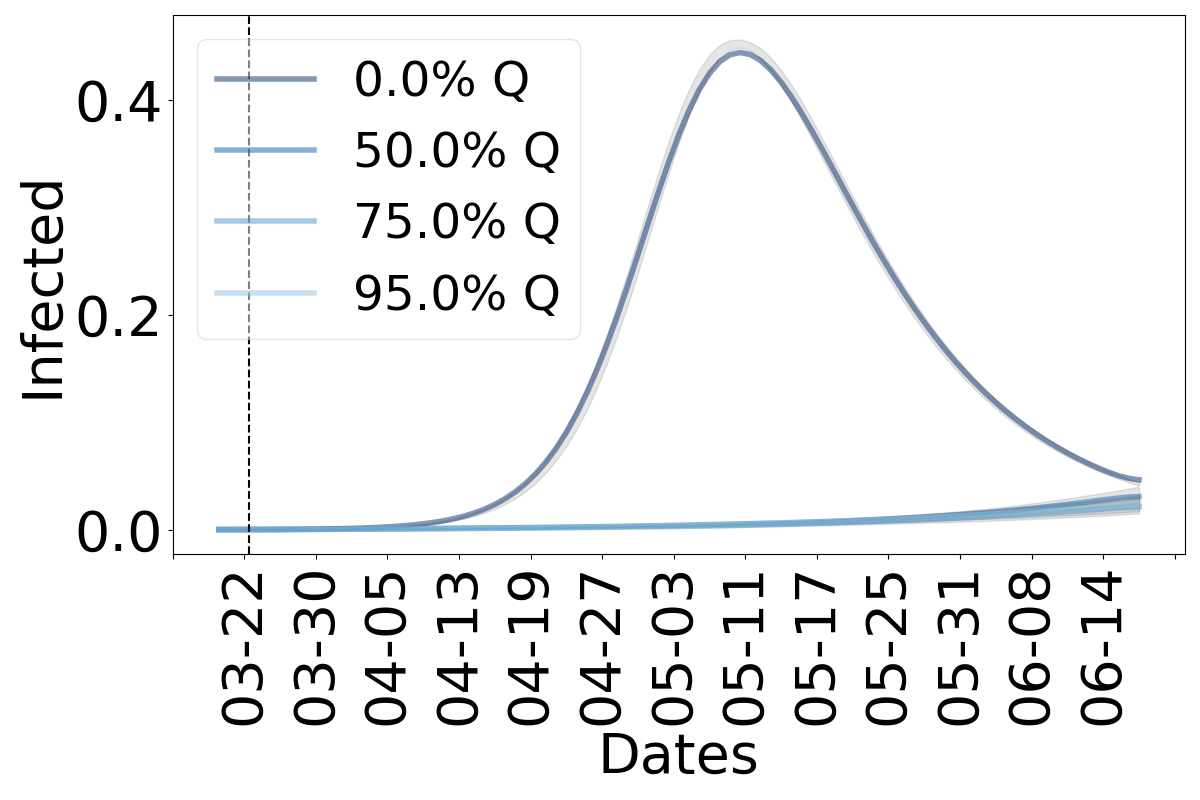}
         \caption{Regular}
         \label{fig:quarantine regular}
     \end{subfigure}
     \hfill
     \begin{subfigure}[b]{0.33\textwidth}
         \centering
         \includegraphics[width=\textwidth]{fig/npi_results/Quarantine_result_ER.png}
         \caption{ER}
         \label{fig:quarantine er}
     \end{subfigure}
     \begin{subfigure}[b]{0.33\textwidth}
         \centering
         \includegraphics[width=\textwidth]{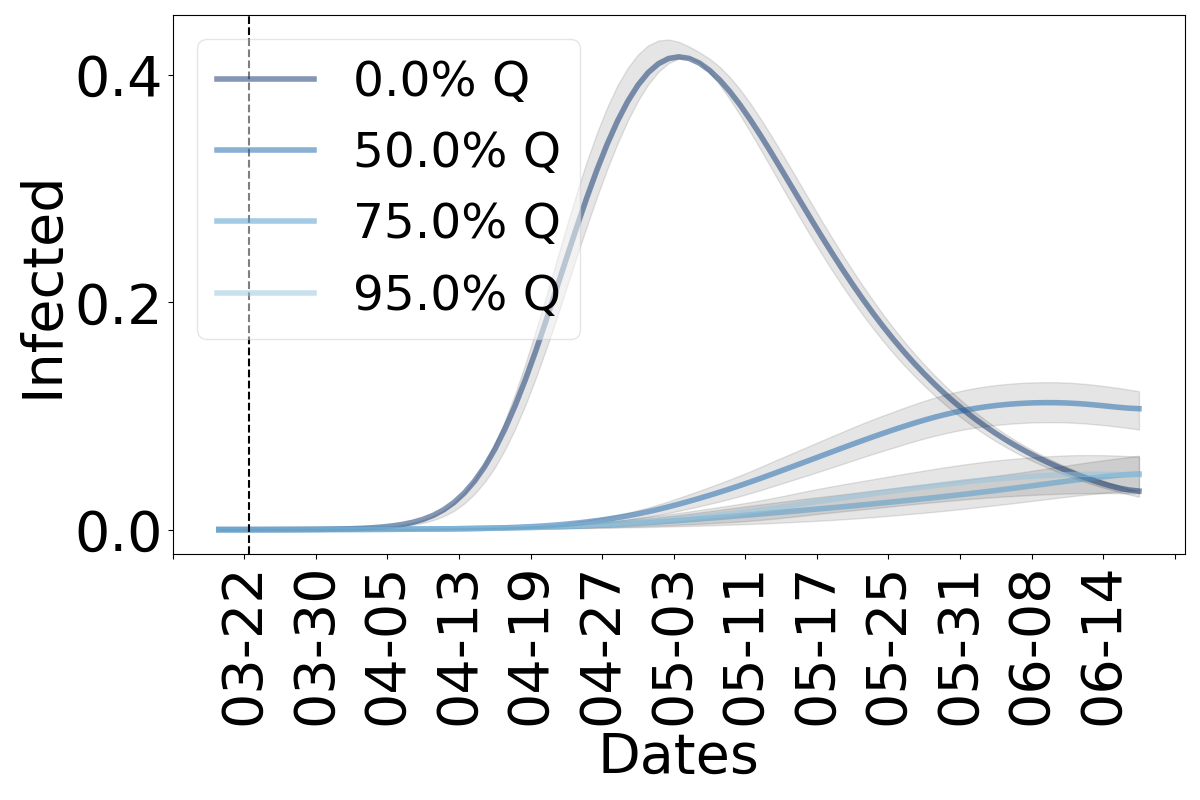}
         \caption{BA}
         \label{fig:quaraintine ba}
     \end{subfigure}
     \hfill
     \begin{subfigure}[b]{0.33\textwidth}
         \centering
         \includegraphics[width=\textwidth]{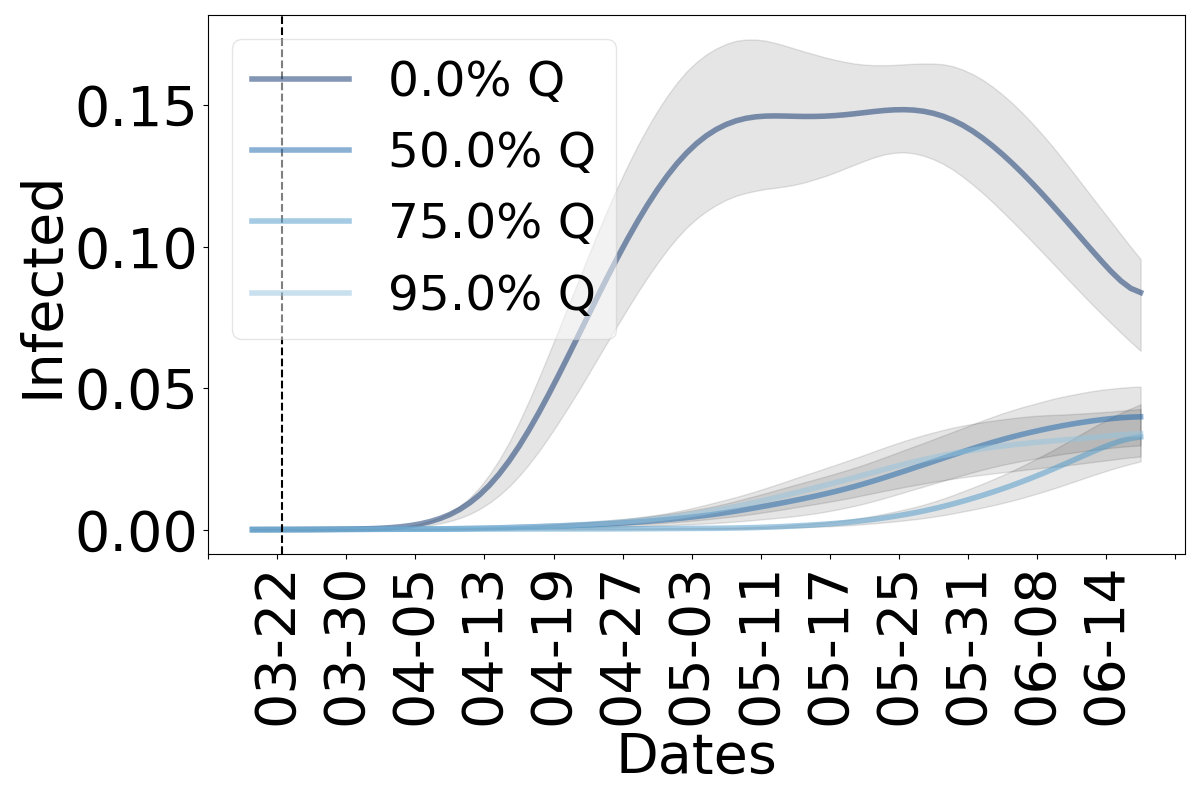}
         \caption{Wifi 1}
         \label{fig:quaraintine wifi 1}
     \end{subfigure}
     \hfill
     \begin{subfigure}[b]{0.33\textwidth}
         \centering
         \includegraphics[width=\textwidth]{fig/npi_results/Quarantine_result_wifi2.png}
         \caption{Wifi 2}
         \label{fig:quarantine wifi 2}
     \end{subfigure}
     \hfill
     \begin{subfigure}[b]{0.33\textwidth}
         \centering
         \includegraphics[width=\textwidth]{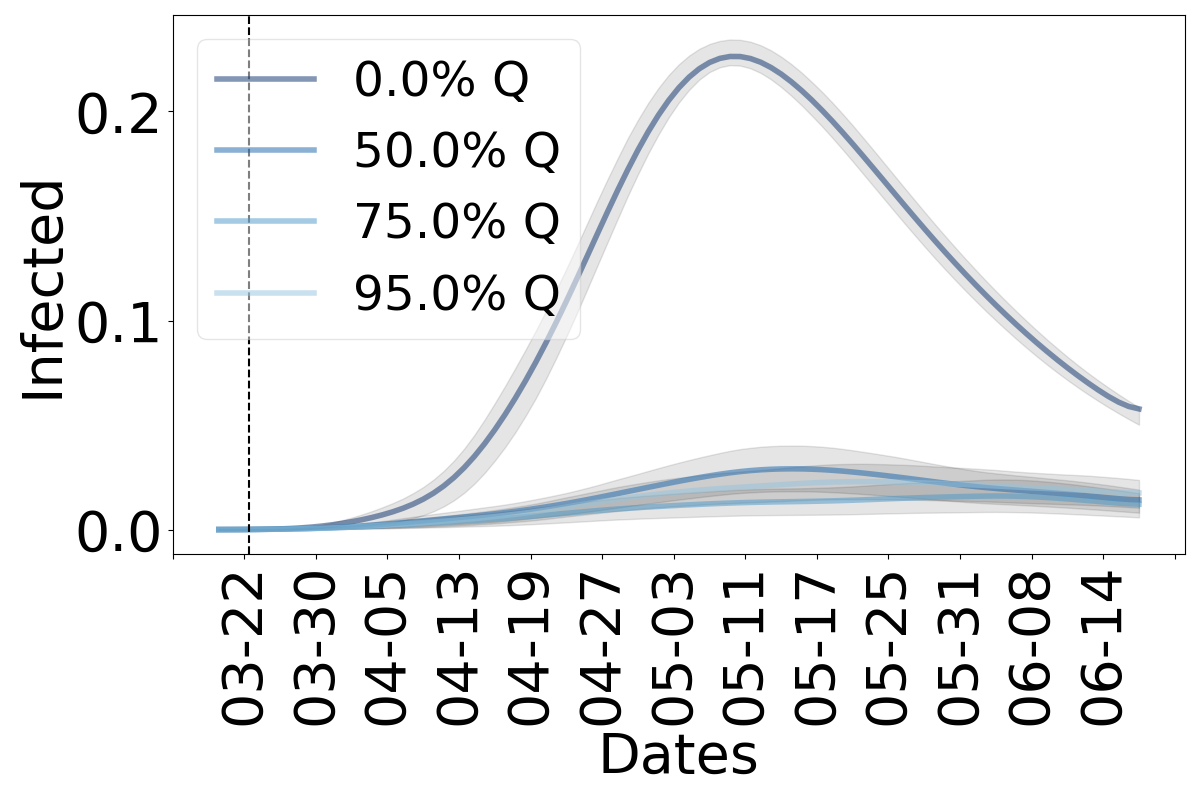}
         \caption{Wifi 3}
         \label{fig:quarantine wifi 3}
     \end{subfigure}
     \hfill
        \caption{Results of quarantining 50, 75, and 95\% of infected and exposed nodes on all graph structures.}
        \label{fig:quarantine}
\end{figure*}

Figure~\ref{fig:quarantine} shows the results of quarantining on all graph structures. Quarantining infected and exposed nodes both reduces and delays the  peak  of  all  infection  curve.  However,  the peak is not delayed as much in the wifi graphs when compared to the regular and ER graphs.

\subsubsection{Social distancing}
\begin{figure*}[h]
     \centering
     \begin{subfigure}[b]{0.33\textwidth}
         \centering
         \includegraphics[width=\textwidth]{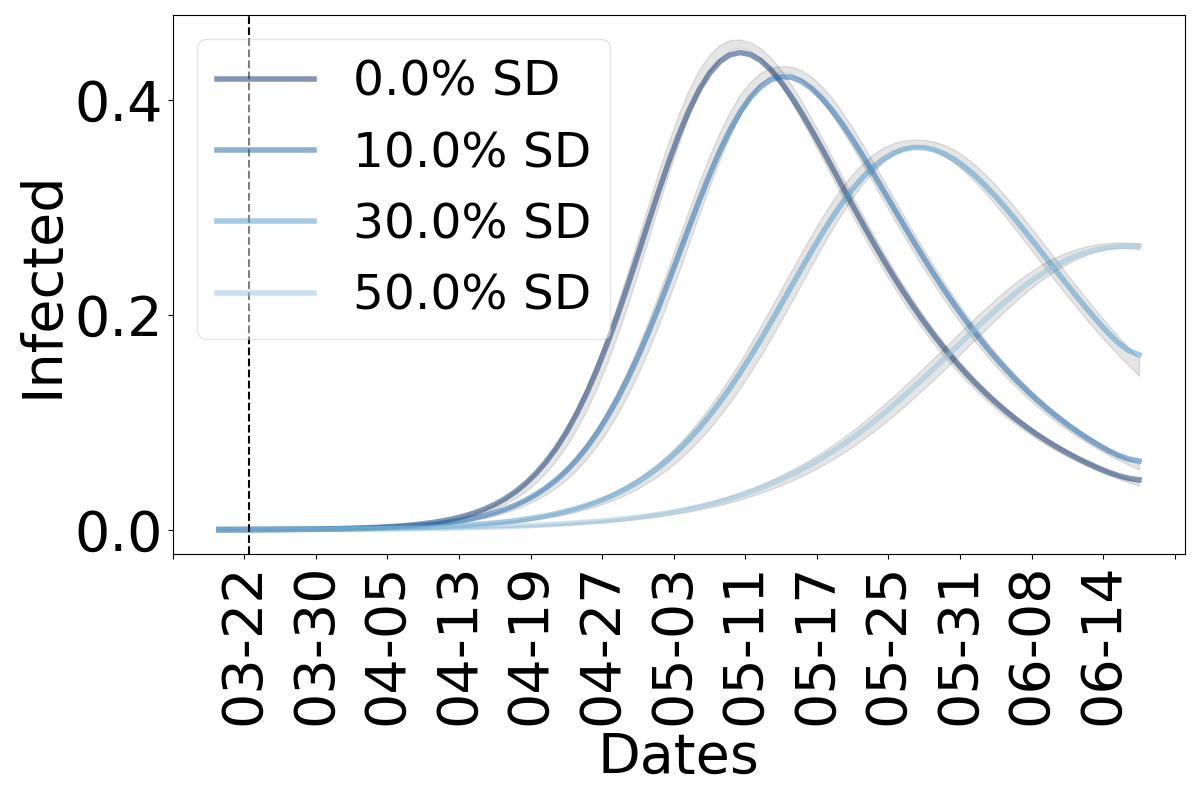}
         \caption{Regular}
         \label{fig:Social distance regular}
     \end{subfigure}
     \hfill
     \begin{subfigure}[b]{0.33\textwidth}
         \centering
         \includegraphics[width=\textwidth]{fig/npi_results/Social_distancing_result_ER.png}
         \caption{ER}
         \label{fig:Social distance er}
     \end{subfigure}
     \begin{subfigure}[b]{0.33\textwidth}
         \centering
         \includegraphics[width=\textwidth]{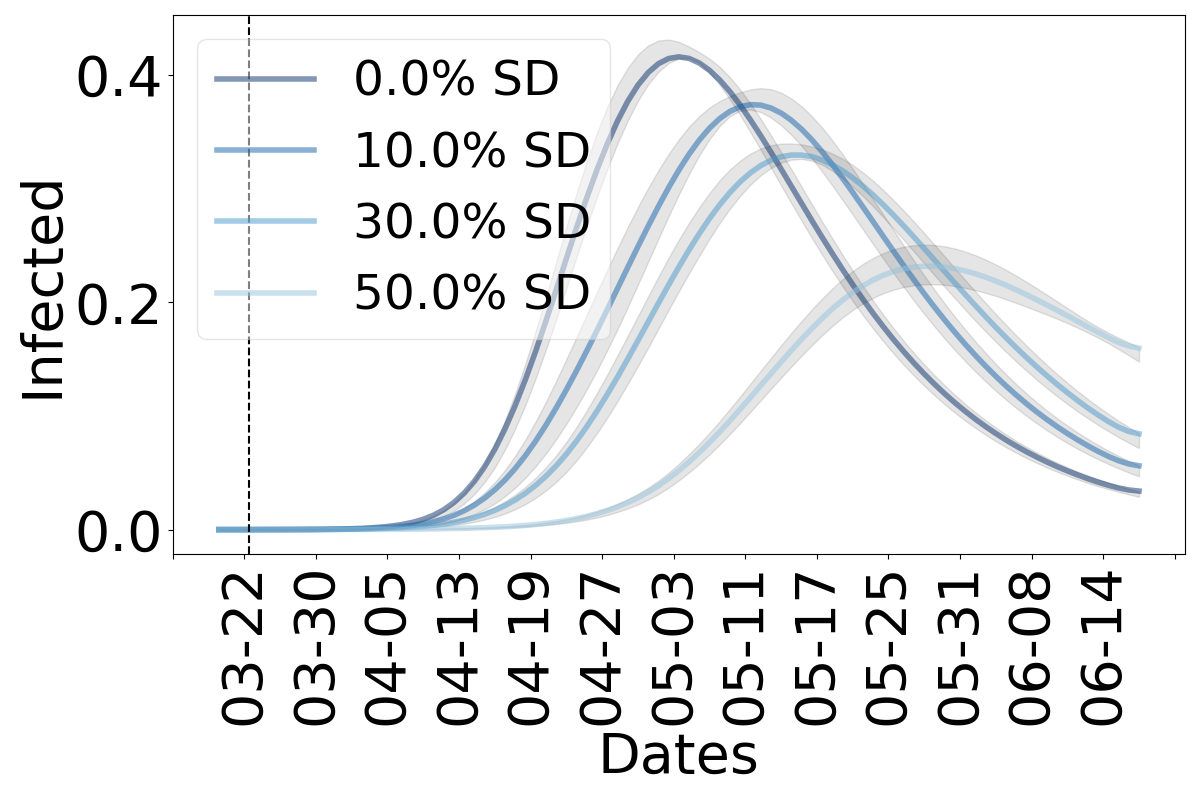}
         \caption{BA}
         \label{fig:social distance ba}
     \end{subfigure}
     \hfill
     \begin{subfigure}[b]{0.33\textwidth}
         \centering
         \includegraphics[width=\textwidth]{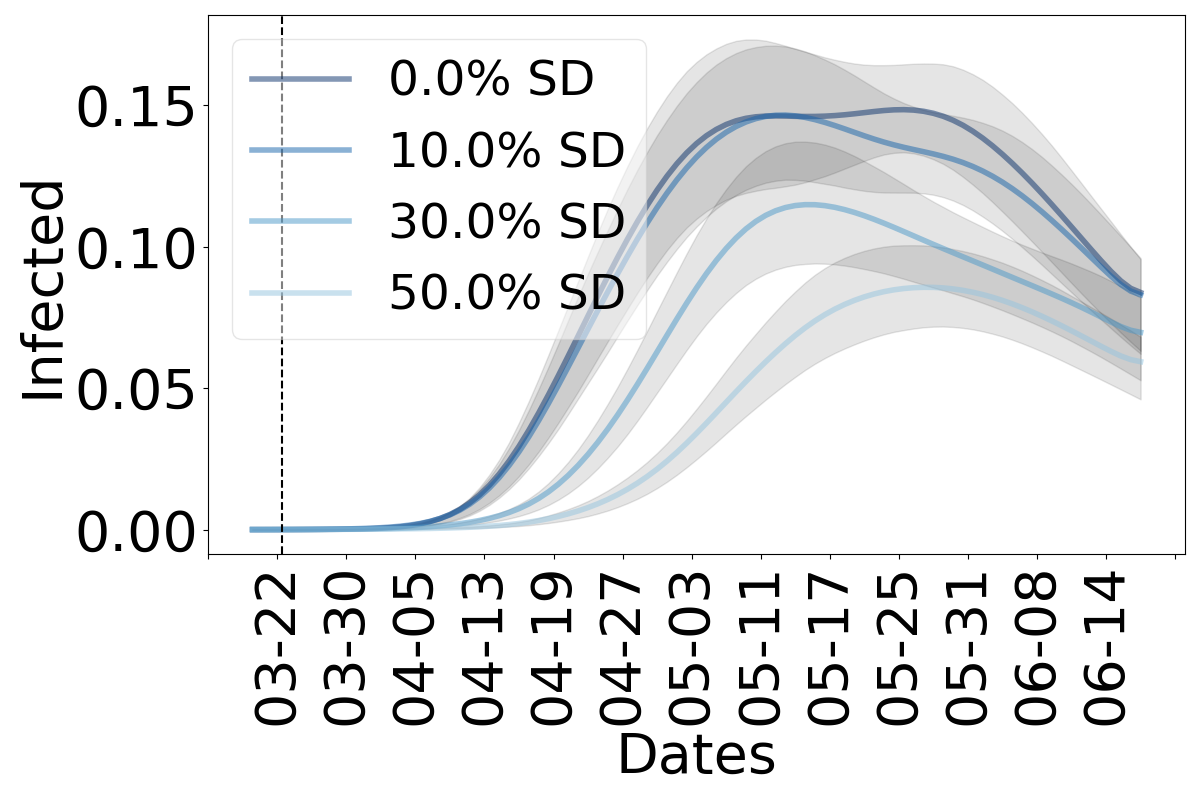}
         \caption{Wifi 1}
         \label{fig:social distance wifi 1}
     \end{subfigure}
     \hfill
     \begin{subfigure}[b]{0.33\textwidth}
         \centering
         \includegraphics[width=\textwidth]{fig/npi_results/Social_distancing_result_wifi2.png}
         \caption{Wifi 2}
         \label{fig:Social distance wifi 2}
     \end{subfigure}
     \hfill
     \begin{subfigure}[b]{0.33\textwidth}
         \centering
         \includegraphics[width=\textwidth]{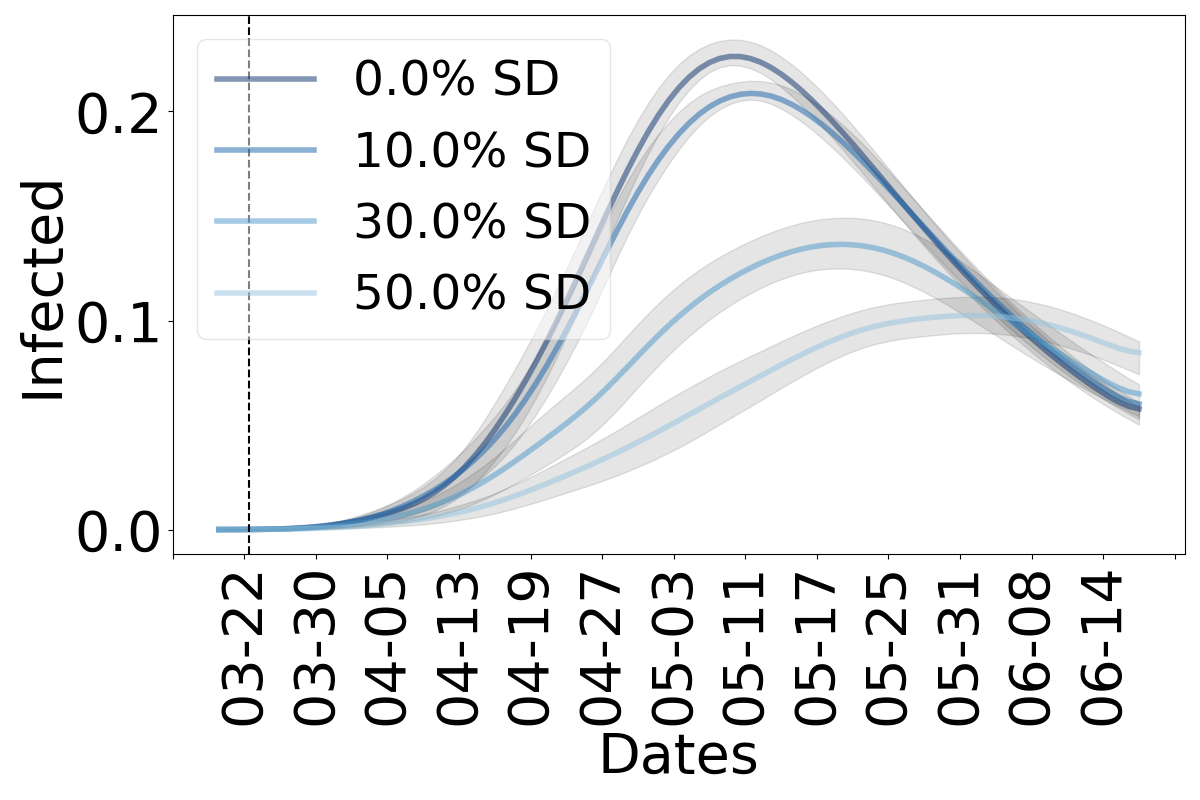}
         \caption{Wifi 3}
         \label{fig:Social distance wifi 3}
     \end{subfigure}
     \hfill
        \caption{Results of applying social distancing by removing 10, 30, and 50\% of edges to all nodes on all graph structures.}
        \label{fig:Social distance}
\end{figure*}

Figure~\ref{fig:Social distance} shows the results of applying social distancing on all networks. Like quarantining, this is effective in reducing the peaks of the infection curve on all networks, but the delay of peaks is only apparent on the synthetic networks.

\subsubsection{Removing hubs}
\begin{figure*}[h]
     \centering
     \begin{subfigure}[b]{0.33\textwidth}
         \centering
         \includegraphics[width=\textwidth]{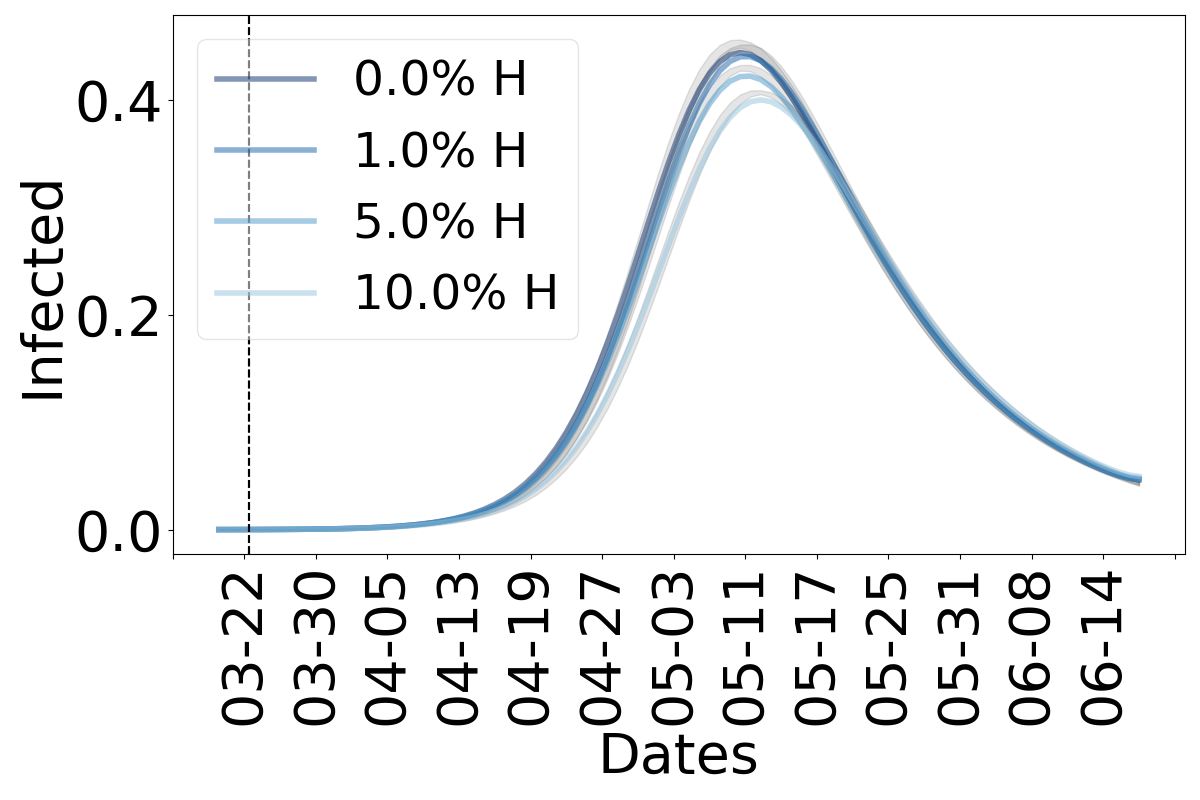}
         \caption{Regular}
         \label{fig:hub regular}
     \end{subfigure}
     \hfill
     \begin{subfigure}[b]{0.33\textwidth}
         \centering
         \includegraphics[width=\textwidth]{fig/npi_results/Remove_Hubs_0.8_result_ER.png}
         \caption{ER}
         \label{fig:hub er}
     \end{subfigure}
     \begin{subfigure}[b]{0.33\textwidth}
         \centering
         \includegraphics[width=\textwidth]{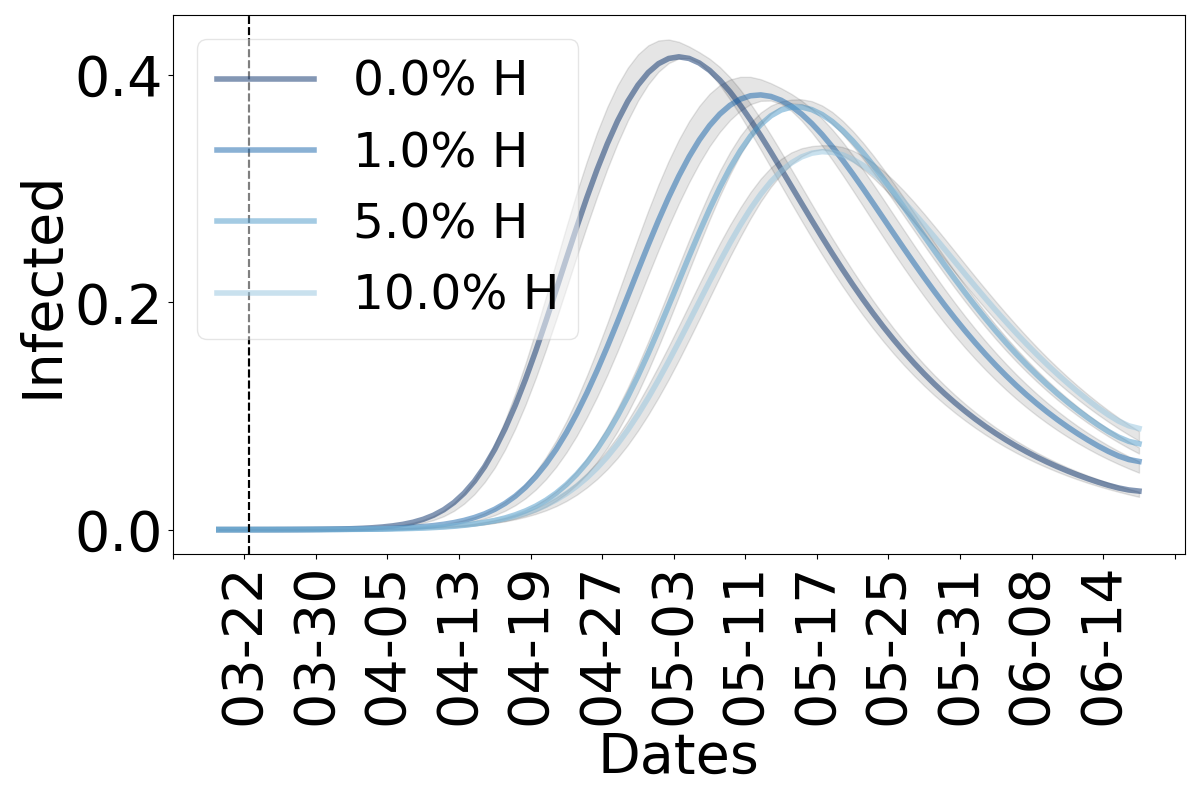}
         \caption{BA}
         \label{fig:hub ba}
     \end{subfigure}
     \hfill
     \begin{subfigure}[b]{0.33\textwidth}
         \centering
         \includegraphics[width=\textwidth]{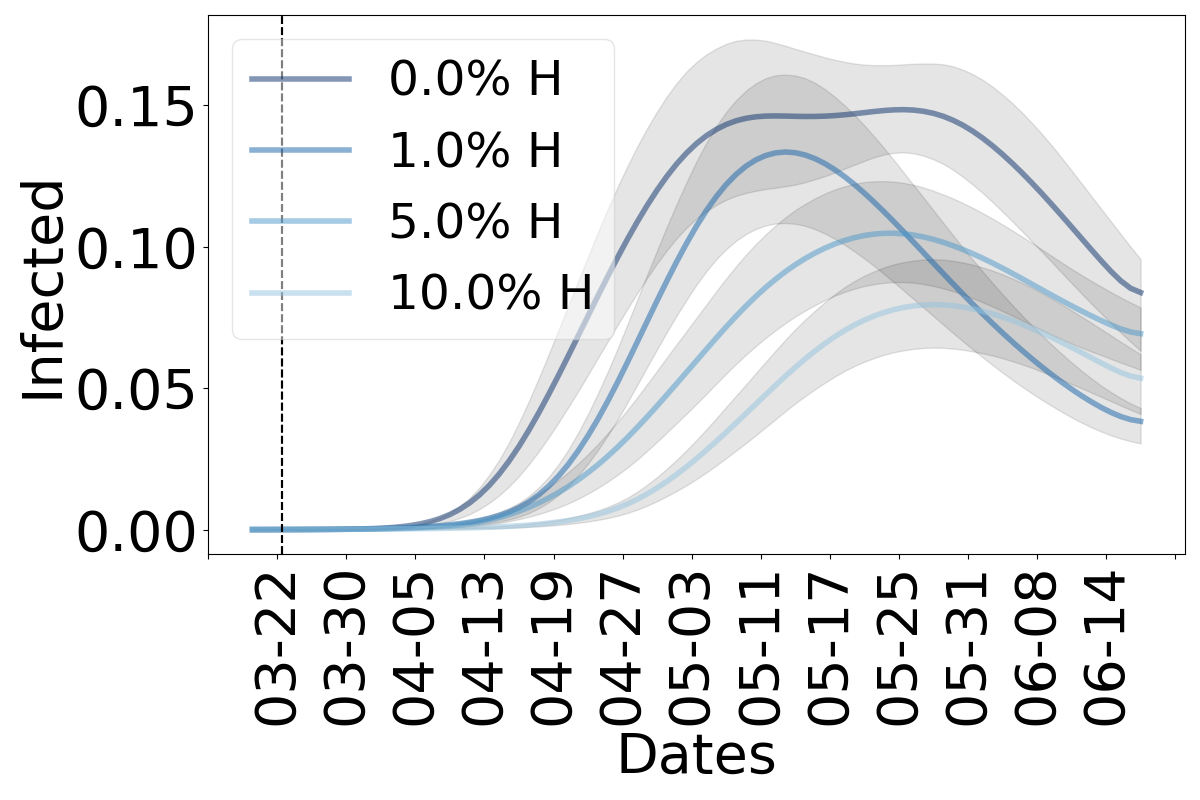}
         \caption{Wifi 1}
         \label{fig:hub wifi 1}
     \end{subfigure}
     \hfill
     \begin{subfigure}[b]{0.33\textwidth}
         \centering
         \includegraphics[width=\textwidth]{fig/npi_results/Remove_Hubs_0.8_result_wifi2.png}
         \caption{Wifi 2}
         \label{fig:hub wifi 2}
     \end{subfigure}
     \hfill
     \begin{subfigure}[b]{0.33\textwidth}
         \centering
         \includegraphics[width=\textwidth]{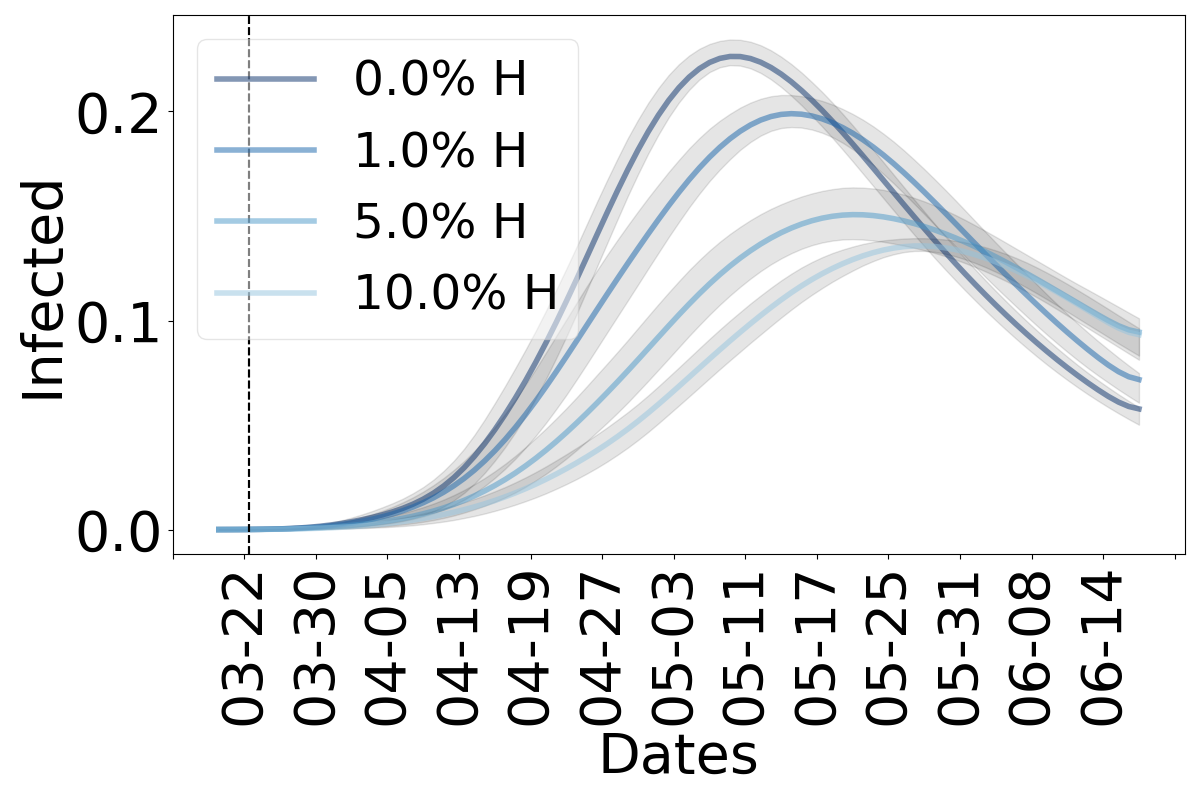}
         \caption{Wifi 3}
         \label{fig:hub wifi 3}
     \end{subfigure}
     \hfill
        \caption{Results of removing 1, 5, and 10\% of hubs from each network. }
        \label{fig:hub}
\end{figure*}
Figure~\ref{fig:hub} shows the results of apply school and business closure on all networks. The ER and regular random networks significantly underestimates the effect of removing hubs.

\subsubsection{wearing masks}
\begin{figure*}[h]
     \centering
     \begin{subfigure}[b]{0.33\textwidth}
         \centering
         \includegraphics[width=\textwidth]{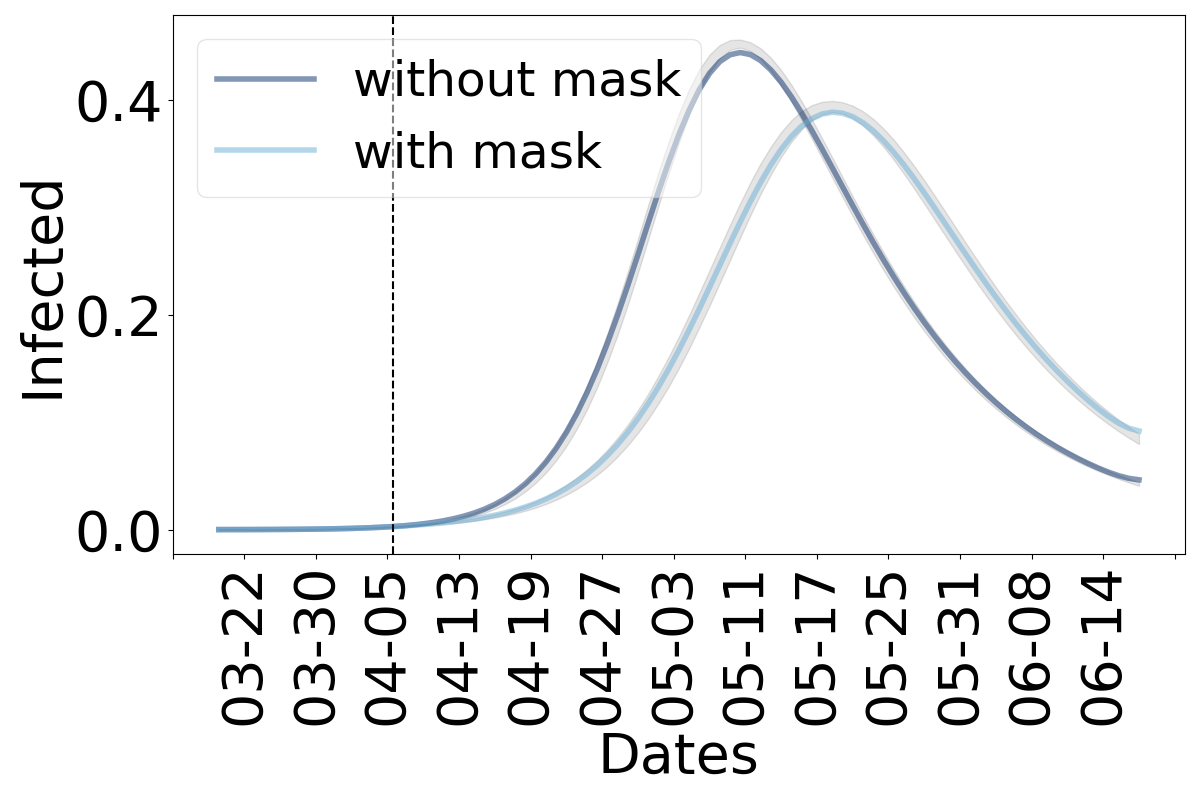}
         \caption{Regular}
     \end{subfigure}
     \hfill
     \begin{subfigure}[b]{0.33\textwidth}
         \centering
         \includegraphics[width=\textwidth]{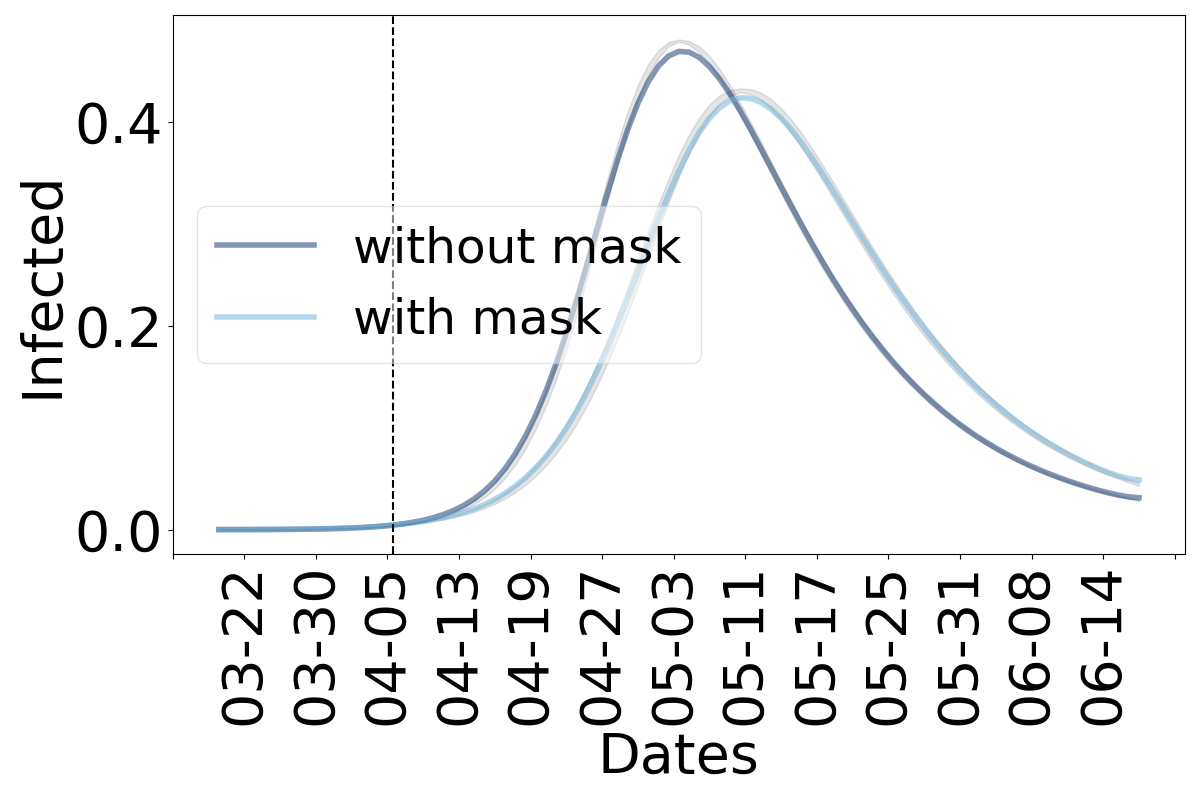}
         \caption{ER}
     \end{subfigure}
     \begin{subfigure}[b]{0.33\textwidth}
         \centering
         \includegraphics[width=\textwidth]{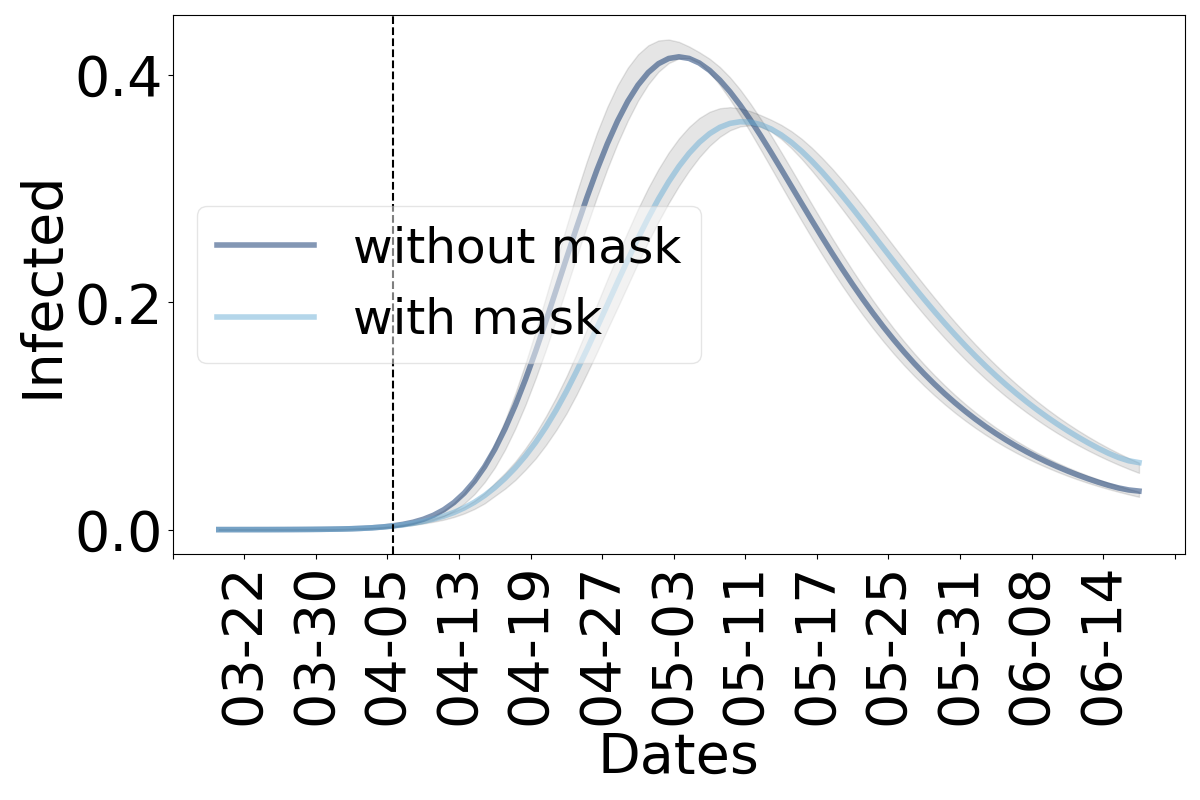}
         \caption{BA}
     \end{subfigure}
     \hfill
     \begin{subfigure}[b]{0.33\textwidth}
         \centering
         \includegraphics[width=\textwidth]{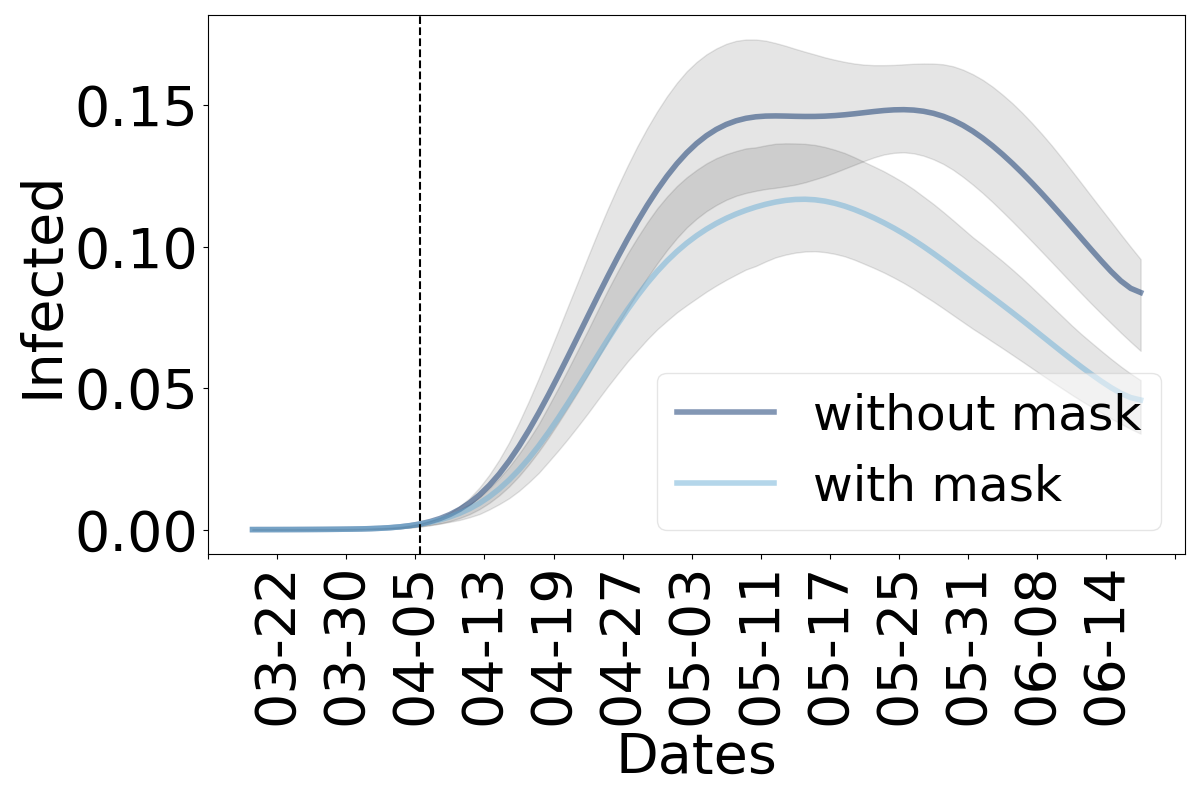}
         \caption{Wifi 1}
     \end{subfigure}
     \hfill
     \begin{subfigure}[b]{0.33\textwidth}
         \centering
         \includegraphics[width=\textwidth]{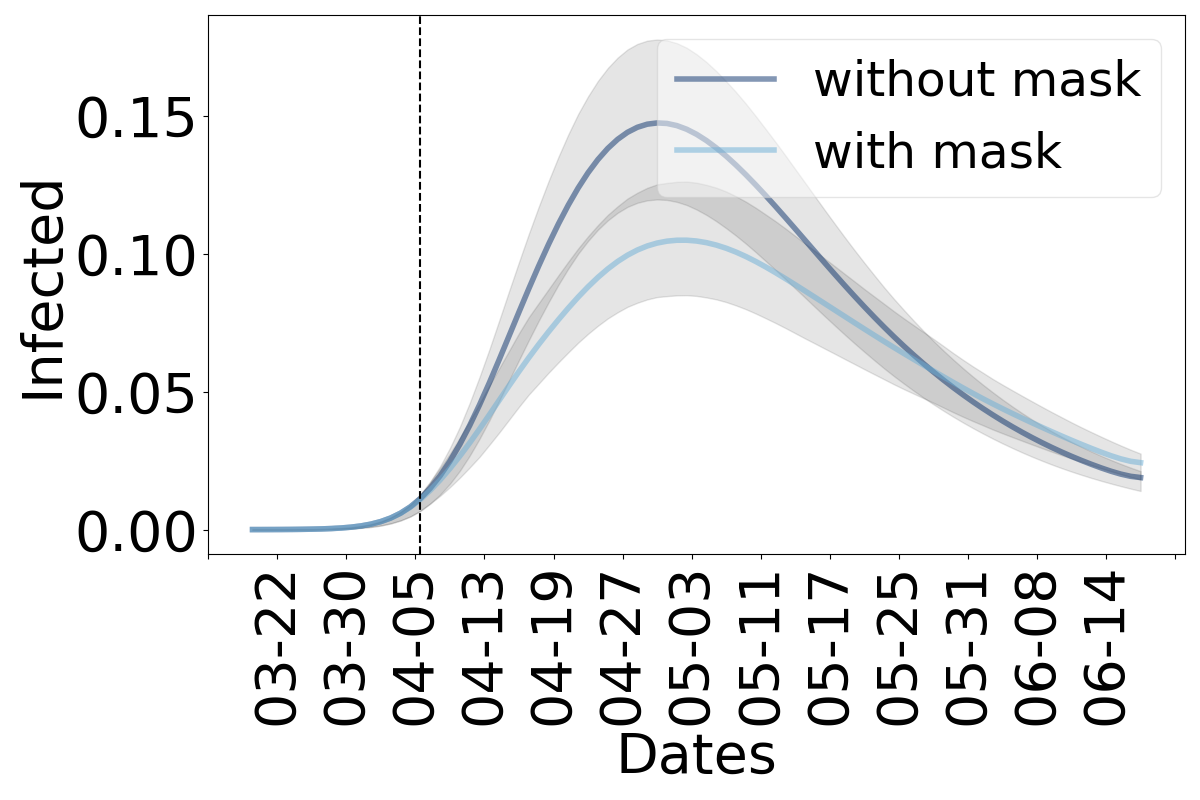}
         \caption{Wifi 2}
     \end{subfigure}
     \hfill
     \begin{subfigure}[b]{0.33\textwidth}
         \centering
         \includegraphics[width=\textwidth]{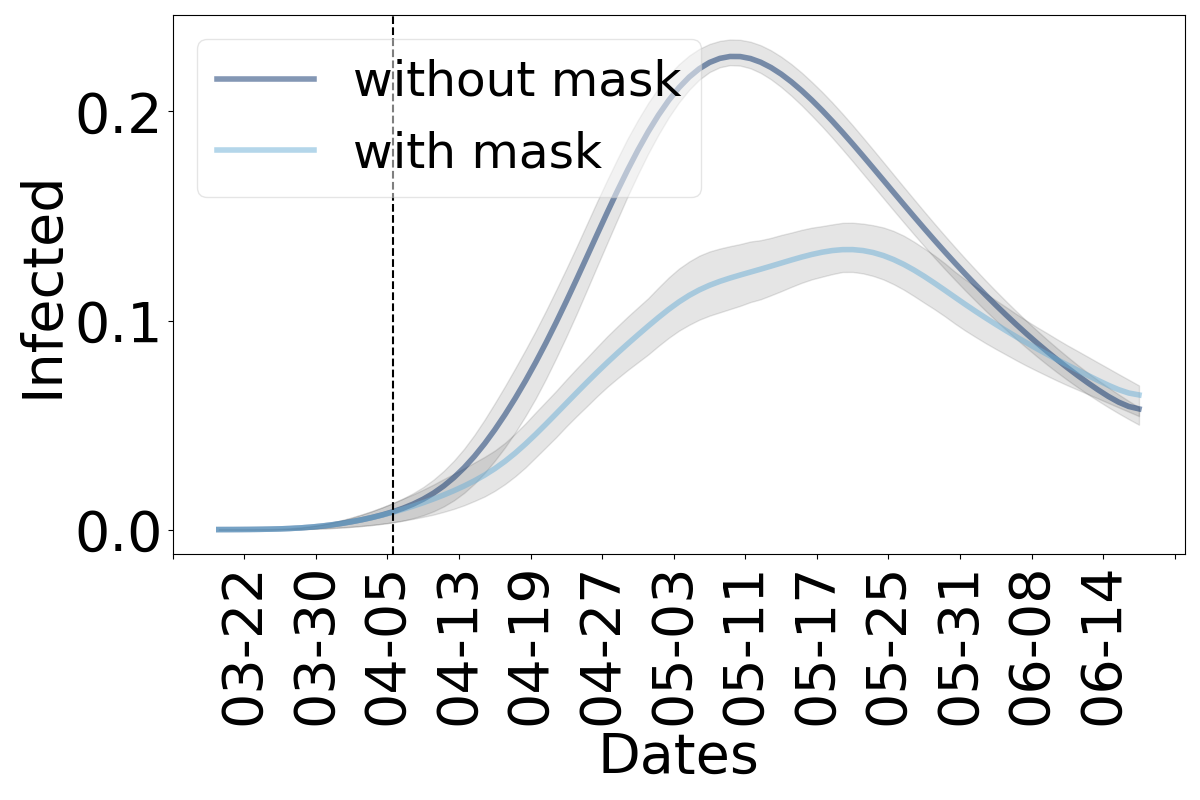}
         \caption{Wifi 3}
     \end{subfigure}
     \hfill
        \caption{Results of with and without masks. }
        \label{fig:mask}
\end{figure*}
Figure~\ref{fig:mask} shows the results of wearing masks and without on each network. 

\subsubsection{All NPIs}
\begin{figure}
    \centering
    \includegraphics[width=0.5\textwidth]{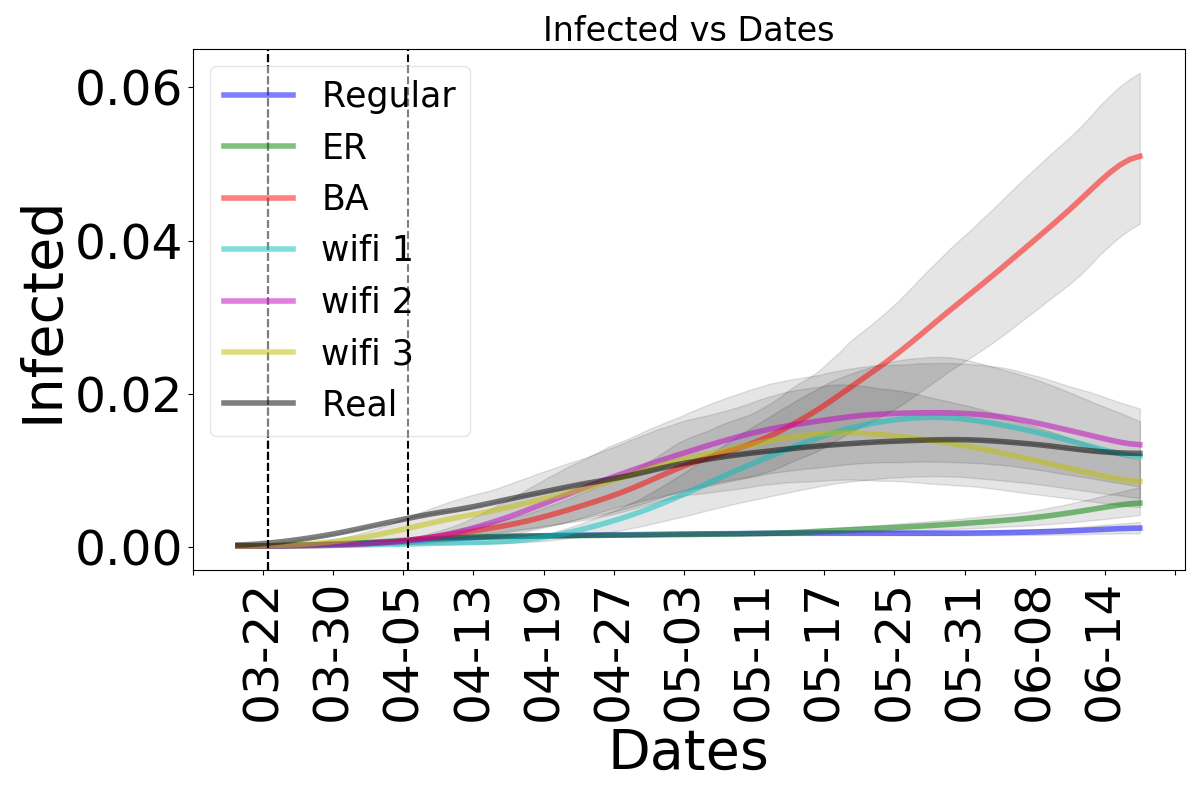}
    \caption{All NPIs applied according to the timeline in Table \ref{tab:NPIs}}
    \label{fig:all}
\end{figure}

Figure~\ref{fig:all} shows the infection curves of all the networks with all NPIs applied. On March 23, 50\% social distancing and 50\% quaranine is applied, and 10\% of hubs are removed with a success rate of 0.8. Wearing mask is applied on April 6. The wifi networks more closely resemble the shape of the real infection curve.

\end{document}